\documentclass[prx,twocolumn,amsmath,amssymb,floatfix,footinbib,bibnotes,longbibliography]{revtex4-2}
\pdfoutput=1

\usepackage[colorlinks=true,citecolor=blue,linkcolor=magenta]{hyperref}
\usepackage{multirow}
\usepackage{amsmath}
\usepackage{bbm}
\usepackage{graphicx}
\usepackage{dsfont}
\usepackage{changepage}
\usepackage{fancyhdr}
\usepackage{amsthm, amssymb}
\usepackage{array}
\usepackage[usenames,dvipsnames]{color}
\usepackage[section]{placeins}
\usepackage[T1]{fontenc}
\usepackage[latin9]{inputenc}
\usepackage[active]{srcltx}
\usepackage{color}
\usepackage{amssymb}
\usepackage{esint}
\usepackage[version=4]{mhchem}

\def \bs{\boldsymbol}

\newcommand{\ci}{\imath}
\newcommand {\apgt} {\ {\raise-.5ex\hbox{$\buildrel>\over\sim$}}\ }
\newcommand {\aplt} {\ {\raise-.5ex\hbox{$\buildrel<\over\sim$}}\ }

\newcommand{\kf}{k_\mathrm{F}}

\def \titlename {Quantum Oscillations in the Magnetization and Density of States of Insulators}
\def \authornames{Animesh Panda$^{1}$, Sumilan Banerjee$^{1}$, Mohit Randeria$^{2}$}
\def \affiliations{$^1$Centre for Condensed Matter Theory, Department of Physics, Indian Institute of Science, Bangalore 560012, India \\
$^2$Department of Physics, The Ohio State University, Columbus, Ohio 43210, USA}
\begin{document}

\title{\titlename}
\author{\authornames}
\affiliation{\affiliations}


\begin{abstract}
 The observation of $1/B$-periodic behavior in Kondo insulators SmB$_6$ and YbB$_{12}$ challenges the conventional wisdom that quantum oscillations (QO) necessarily arise from Fermi surfaces in metals. We revisit recently proposed theories for this phenomena, focusing on a minimal model of an insulator
 with a hybridization gap between two opposite-parity light and heavy mass bands with an inverted band structure. We show that there are characteristic differences between the QO frequencies in the magnetization and the low-energy density of states (LE-DOS) of these insulators, in marked contrast with metals where {\it all} observables exhibit oscillations at the {\it same} frequency. The magnetization oscillations are shown to arise from all occupied Landau levels and exhibit the same frequency as the unhybridized case. The LE-DOS oscillations arise from gap-edge states in a disorder-free system and exhibit a beat pattern between two distinct frequencies at low temperature. Disorder induced in-gap states lead to an additional contribution to the DOS at the unhybridized frequency. The temperature dependence of the magnetization and DOS oscillations are qualitatively different and both show marked deviations from the Lifshitz-Kosevich form. We also compute transport to ensure that we are probing a regime with insulating upturns in the dc resistivity.
\end{abstract}

\maketitle

\section{Introduction}
Quantum oscillations~\cite{shoenberg} have long been considered to be {\em the} most direct probe of the Fermi surface in metals, the locus of gapless electronic excitations in ${\bs k}$-space. It thus came as a great surprise that the Kondo insulators SmB$_6$ and YbB$_{12}$ exhibit $1/B$-periodic oscillations~\cite{sebastian_2015,Hartstein2018,Sebastian_2018,Li_Matsuda_2018}, even though these materials do not have any gapless electronic excitations in the bulk. Soon after, it was pointed out by Knolle and Cooper (KC) \cite{KC_2015} that a simple model of an insulator with a hybridization gap exhibits deHass van Alphen (dHvA) oscillations in the magnetization even in the absence of a Fermi surface. Though the KC theory does not capture the observed $T$-dependence of the oscillation amplitude in SmB$_6$, it is very important from a conceptual point of view. 

Predictions of the KC model~\cite{KC_2017_1} led to quantum oscillation experiments in semiconductors: InAs/GaSb quantum wells~\cite{RRDu_2019,Samarth_2019}. The KC ideas have been extended to include more realistic hybridization~\cite{FW_2016} and impurity states~\cite{LF_2018} with a focus on the low-energy density of states (LE-DOS) oscillations, which are a proxy for the Shubnikov deHass (SdH) oscillations in transport. The Kondo insulators are strongly correlated systems, and many exotic mechanisms (involving Majorana fermions, or fractionalized phases, or topological excitations or magnetoexcitons) \cite{Baskaran2015,Erten2017,Sodemann2018,Chowdhury2018,Varma2020,Knolle2017} have also been proposed for understanding the observed quantum oscillations. It is fair to say that no existing theory has been able to fully account for all of the features observed in the data.

\begin{table}
\centering
\begin{tabular}{|p{1.6cm}|p{1.4cm}|p{1.5cm}|p{1.4cm}|p{1.8cm}|}
\hline
 Observable & States that contribute & Frequency & Dingle damping & Temperature dependence, $T\to 0$ limit \\
\hline
\multirow{2}{*}{LE-DOS} & Gap-edge states & $F_0\pm \delta F$ & \hspace{1mm} $e^{-\frac{\pi\Delta_IT}{(\hbar\omega_c)^2}}$& non-LK, $e^{-\frac{\Delta_I}{2T}}$ \\
 & In-gap states & $F_0$ & \hspace{1mm}$e^{-\frac{\pi\Delta_I}{2\hbar \omega_c}}$ & LK-like, constant \\ \hline  
$M$ &  All states below $\mu$& $F_0$ & \hspace{1mm}$e^{-\frac{\pi\Delta_I}{2\hbar \omega_c}}$ & non-LK, constant\\
\hline
\end{tabular}
\caption{{\bf Summary of results:} The principal characteristics of low energy DOS and magnetization ($M$) oscillations in a hybridization-gap insulator 
with an indirect gap $\Delta_I \gg$ the impurity broadening. $F_0=(\hbar/2\pi e)\pi\kf^2$ is the unhybridized frequency,  $\delta F=(m_2-m_1)\Delta_I/(4\hbar e)$, and 
the cyclotron frequency $\omega_c=eB/(m_1+m_2)$.}
\label{tab:Summary}
\end{table}

In this paper we revisit the hybridization gap insulator~\cite{KC_2015,FW_2016,LF_2018,Pal2016,Pal2017,Grubinskas2018} and ask the following question: what are the characteristic 
differences between the quantum oscillations in such an insulator and those that arise from a Fermi surface in a metal?
Our answers can be stated simply. In the insulator, magnetization oscillations (dHvA) are governed by all of the occupied Landau levels,
and the oscillation frequency is governed by the area of the Fermi surface that would have existed in the absence of any hybridization.
On the other hand, the SdH oscillations in the LE-DOS are dominated by gap-edge states and exhibit a beat pattern between two distinct frequencies, on either side of the unhybridized frequency, at low but non-zero temperatures in the disorder-free insulator. Disorder induces states within the gap states and this leads to an additional oscillation at the unhybridized frequency. These results are qualitatively different from metals, where all oscillations -- dHvA and SdH -- occur at the same frequency given by the extremal Fermi surface area. In addition, there are also differences between the $T$-dependences of the oscillation amplitude in magnetization and in LE-DOS, neither of which shows the standard Lifshitz-Kosevich (LK) form well known in metals~\cite{shoenberg}, as well as in the Dingle factors.

\begin{figure*}
	\centering
	{\includegraphics[width=0.9\textwidth]{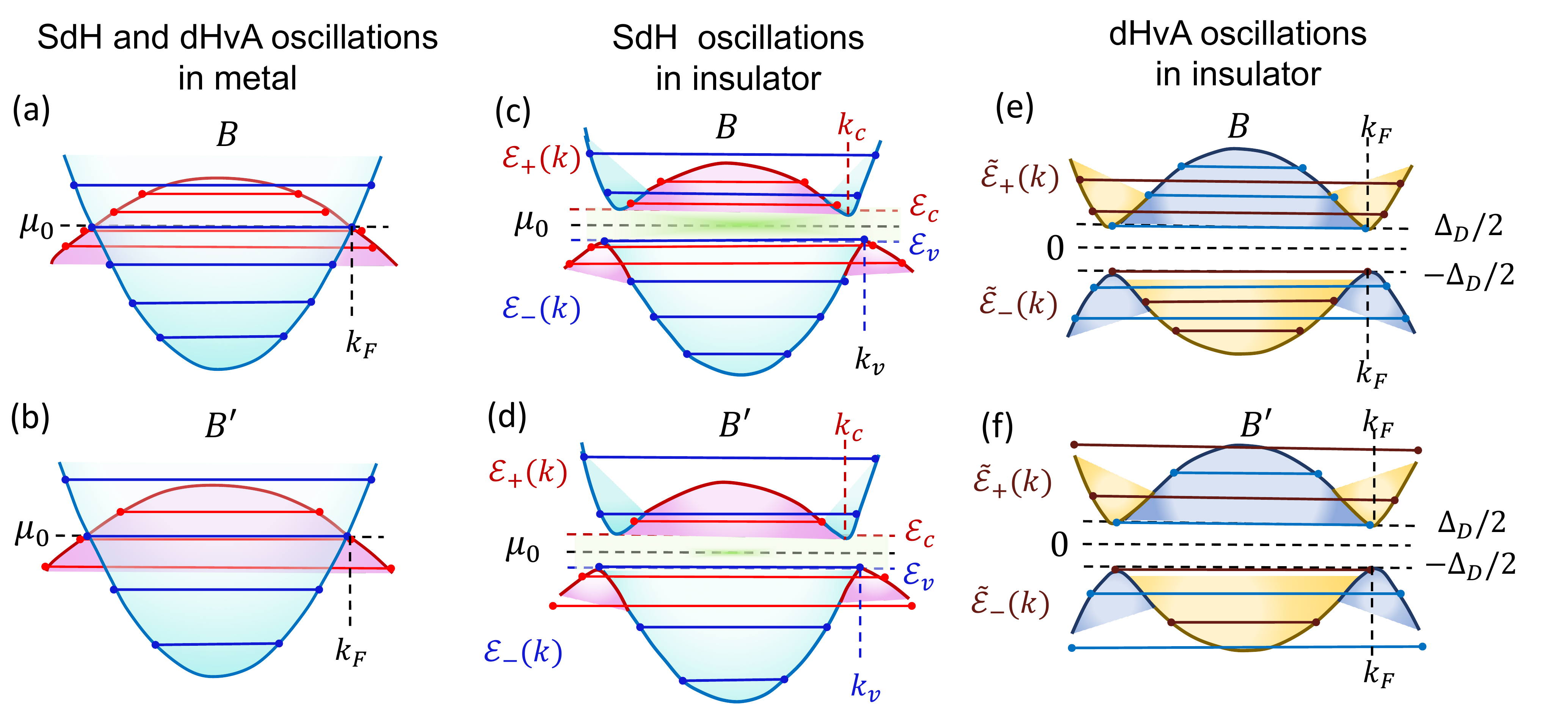}}
	\caption{
	{\bf Physical origin of the distinct frequencies for LE-DOS and Magnetization oscillations:}
	(a), (b) Band structure and Landau levels in the limit of zero hybridization $v=0$. For chemical potential at the band crossing energy $\mu_0=\hbar^2\kf^2/2m_1$ at the wavevector $\kf$, the SdH and dHvA oscillations have the same frequency $F_0$ determined by the Fermi surface area $\pi\kf^2$. The fields $B$ and $B'$ ($B<B'$) correspond to two successive crossing of an LL through $\mu_0$. (c), (d) At finite hybridization $v\neq 0$, the LE-DOS or SdH oscillations arise through thermal activation from crossing of energy levels through the hybridization-gap edges $\mathcal{E}_c$ and $\mathcal{E}_v$ of the conduction [$\mathcal{E}_+(k)$] and valence [$\mathcal{E}_-(k)$] band, respectively. The areas $\pi k_c^2$ and $\pi k_v^2$ at the gap edges determine the frequencies. In addition, in the presence of disorder, the impurity-induced in-gap DOS (green shaded) has $1/B$-periodic modulation with frequency $F_0$. (e), (f) The dHvA oscillations in magnetization arise from a \emph{fictitious} particle-hole symmetric band structure $\tilde{\mathcal{E}}_{\pm}(k)$ centered around zero energy. At $T=0$, the oscillations occur due to sequential entries or exits of additional energy levels, e.g. at fields $B$ and $B'$, into the \emph{electron-like} part (yellow shaded) of $\mathcal{E}_-(k)$ from the hole-like part (blue shaded) of the band through the gap edge at $-\Delta_D/2$. The latter corresponds to the semiclassical orbit at wavevector $\kf$ with area $\pi\kf^2$, and thus the frequency $F_0$ of dHvA oscillations in the hybridization-gap insulator.  
	}
	\label{fig:schematic} 
\end{figure*}

Our main results are summarized in Table 1. These are obtained using analytical calculations in the semi-classical regime, which use saddle point methods, together with extensive numerical calculations, and give insight into the frequency, phase, and amplitude of the quantum oscillations and their dependence on temperature, magnetic field  and disorder. We introduce in Section II our minimal model of a hybridization gap insulator and describe its Landau level spectrum; see Fig.~\ref{fig:Model}. 
In Section III we explain the physical origin of the differences between the dHvA and SdH oscillations in an insulator, which are summarized in Fig.~\ref{fig:schematic}.
Our analytical and numerical results for the low energy DOS are described in Section IV and the results for magnetization in Section V. We conclude in Section VI with 
a brief discussion of quantum oscillation experiments in Kondo insulators and semiconductor quantum wells. Additional details of the analytical and numerical calculations are provided in the appendices.
 
\section{Model} \label{sec:Model}

\begin{figure}[htb!]
	\centering
	{
		\includegraphics[width=0.4\textwidth]{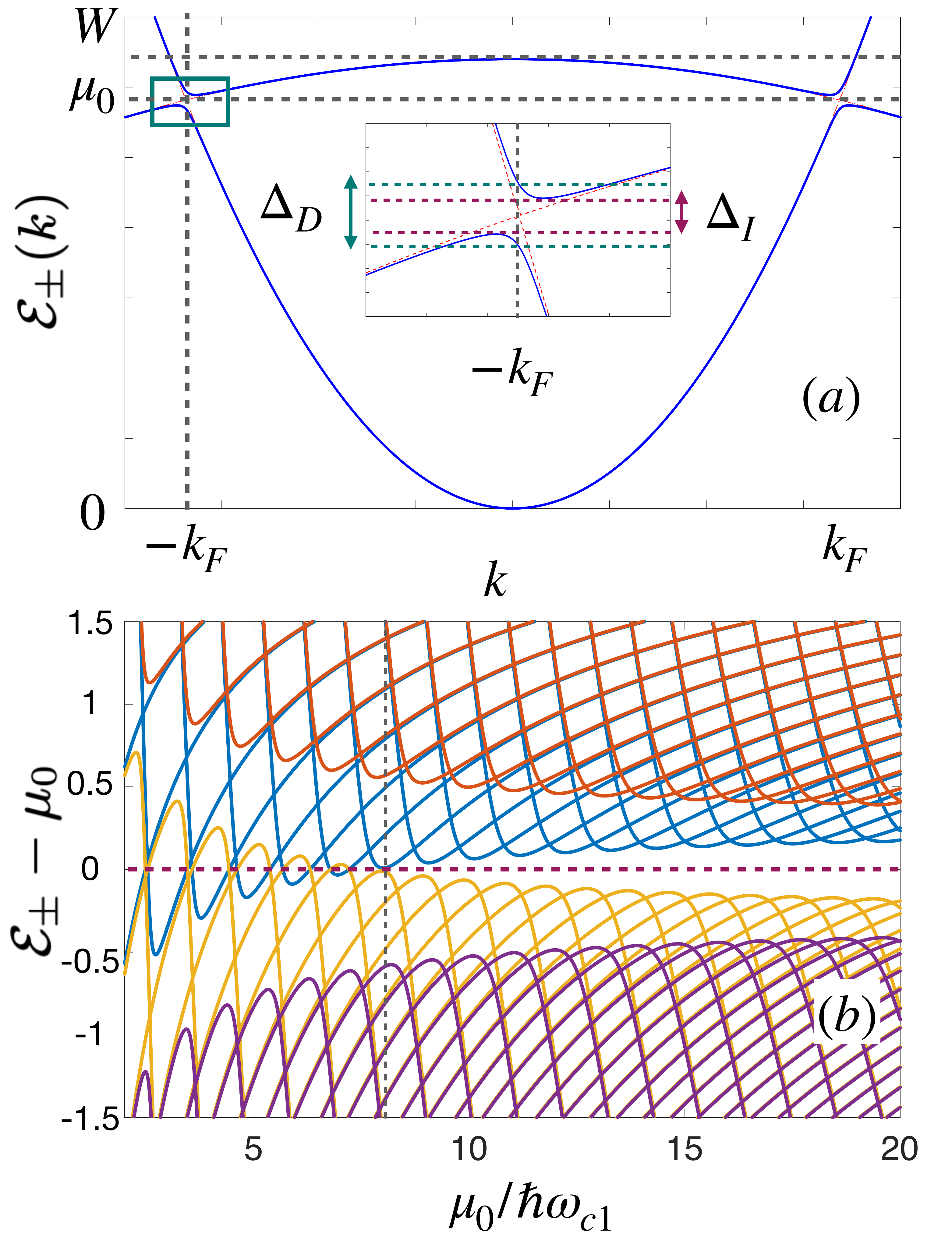}
	}
	
\caption{{\bf Band structure and Landau levels:} (a) Energy dispersion $\mathcal{E}_{\pm}(k)$ in the absence of disorder. The inset
shows the indirect ($\Delta_I$) and direct ($\Delta_D$) gaps. 
(b) Energy levels $\mathcal{E}_{l,b\pm}$ [Eq.\eqref{eq:LLeigen}] for different LL indices plotted as a function of 
$\mu_0/\hbar\omega_{c1}\propto 1/B$. We focus here on the regime $B < B_c$, the critical field (vertical dashed line) above 
which  the system undergoes a field-induced insulator to metal transition;  see text and Appendix \ref{sec:S_Bc} for details. Energy is given in units of $\Delta_D$.
	}
	\label{fig:Model} 
\end{figure}

We consider a two-dimensional (2D) model of an insulator with two opposite-parity bands, a light `$d$' band and an inverted heavy `$f$' band [see Fig.\ref{fig:Model}(a)], with $p$-wave hybridization, described by the Hamiltonian $\mathcal{H}=\sum_{\bs{k}}(d_{\bs{k}}^\dagger~~f_{\bs{k}}^\dagger)H_0(\bs{k})(d_{\bs{k}}~~f_{\bs{k}})^T$, where,
\begin{equation}
H_0(\bs{k})=\begin{bmatrix}
\epsilon_1(\bs{k})\mathbbm{1}  & v\bs{k}\cdot \bs{\sigma}\\v\bs{k}\cdot \bs{\sigma} & \epsilon_2(\bs{k})\mathbbm{1} \label{eq:Model}
\end{bmatrix}, 
\end{equation}
Here $\bs{k}=(k_x,k_y)$, $\bs{\sigma}=(\sigma_x,\sigma_y)$ are Pauli matrices and $\mathbbm{1}$ is the identity matrix in the spin space for the electron 
operators $d_{\bs{k}}=(d_{\bs{k}\uparrow}~~d_{\bs{k}\downarrow})^T$ and $f_{\bs{k}}=(f_{\bs{k}\uparrow}~~f_{\bs{k}\downarrow})^T$.
The dispersion of the unhybridized bands is $\epsilon_1(\bs{k})=\hbar^2k^2/2m_1$ and $\epsilon_2(\bs{k})=W-\hbar^2k^2/2m_2$. 
Unless otherwise mentioned, we set the chemical potential $\mu$ at $\mu_0=Wm_+/m_1$, the energy corresponding to the crossing of the unhybridized bands, where $m_\pm=m_1m_2/(m_2\pm m_1)$. The Fermi wave vector $\kf=\sqrt{2m_+W}/\hbar$ is determined by $W$ the maximum of the $f$ band. 

The hybridization only couples spin $\uparrow$ ($\downarrow$) in the first band with spin $\downarrow$ ($\uparrow$) in the second. The parameter $v$ controls the hybridization gap. 
As shown in Fig.\ref{fig:Model}(a), the insulator has a direct band gap $\Delta_D=2\sqrt{2m_+W}v/\hbar$ and an 
indirect band gap $\Delta_I=2[\sqrt{m_1m_2}/(m_1+m_2)]\Delta_D$. 
We choose $v$ so that $\Delta_D\ll \mu_0$, so that the hierarchy of energy scales is $m_+ v^2/\hbar^2\ll\Delta_I<\Delta_D\ll\mu_0<W$. 

The minimal model of Eq.\eqref{eq:Model} has been widely used to study electronic properties ~\cite{Onur_2015} and quantum oscillations ~\cite{FW_2016,LF_2018} in Kondo insulators. It also has close similarity with models of \ce{InAs/GaSb} quantum wells ~\cite{KC_2017_1,RRDu_2019,Samarth_2019}.

We incorporate the effects of impurities, following ref.~\cite{LF_2018}, with an effective non-Hermitian Hamiltonian $H(\bs{k})$
obtained by replacing $\epsilon_j (\bs{k}) \to \epsilon_j(\bs{k})-\ci\Gamma_j$ for bands $j=1,2$ in Eq.\eqref{eq:Model}.
The frequency- and momentum-independent imaginary self energies are impurity scattering rates
with~\cite{LF_2018} $\Gamma_1 > \Gamma_2 \geq 0$ when $m_1 < m_2$.

$H(\bs{k})$ can be diagonalized to obtain complex eigenvalues $\mathcal{E}_\pm(k)=( \epsilon_1+\epsilon_2-\imath \Gamma 
    \pm \sqrt{(\epsilon_1-\epsilon_2-\imath \gamma)^2+4v^2k^2})/2$,
where $\Gamma=\Gamma_1+\Gamma_2$ and $\gamma=\Gamma_1-\Gamma_2$.
Each eigenvalue is two-fold degenerate given the $\uparrow\downarrow$ and $\downarrow\uparrow$ hybridization. 

Disorder leads to a finite DOS at the chemical potential. However, we can still distinguish a \emph{semimetallic} ($|\gamma|\geq \Delta_D$) regime with zero gap and an \emph{insulating} ($|\gamma|<\Delta_D$) regime with a finite gap ${\rm Re}[\mathcal{E}_+(\kf)-\mathcal{E}_-(\kf)]=\sqrt{\Delta_D^2-\gamma^2}$ at $\kf$~\cite{LF_2018}, in the quasiparticle energy defined by the real part of the complex eigenvalues~\cite{LF_2018}. 
We focus here on the insulating regime; the semimetallic regime gives rise to standard quantum oscillations like a metal.
 
The effect of Landau quantization in the presence of a magnetic field ${\bf B} = B\hat{\bf z}$ in the Hamiltonian $H(\bs{k})$
breaks the degeneracy of the eigenvalues for the $\uparrow\downarrow$ and $\downarrow\uparrow$ combinations, and we get four eigenvalues
\begin{align}
\mathcal{E}_{l,b\pm}=\frac{\epsilon_{1,\ell_b}+\epsilon_{2,\ell'_b}-\ci\Gamma\pm\sqrt{(\epsilon_{1,\ell_b}-\epsilon_{2,\ell'_b}-\ci\gamma)^2+\frac{8lv^2eB}{\hbar}}}{2} .\label{eq:LLeigen} 
\end{align}
Here the Landau level (LL) index $l\geq 1$ with $\ell_b=l,\ell'_b=l-1$ for $b=\uparrow\downarrow$, and $\ell_b=l-1,\ell'_b=l$ for $b=\downarrow\uparrow$ hybridizations.
The $\pm$ signs refer to antibonding/bonding bands. $\epsilon_{1,l}=\hbar\omega_{c1}(l+1/2)$ and $\epsilon_{2,l}=W-\hbar\omega_{c2}(l+1/2)$ are LL energies for the unhybridized bands with cyclotron frequencies $\omega_{c1}=eB/m_1$ and $\omega_{c2}=eB/m_2$.
The $l=0$ LLs remain unchanged with energies $\epsilon_{1,0}$ and $\epsilon_{2,0}$ even for non-zero hybridization, but
these are not relevant for the semiclassical limit $\mu_0/\hbar\omega_{c1}\gg 1$ that we focus on. 

In the semiclassical limit $\ell'\approx\ell=l$ and $8lv^2eB/\hbar\approx 8l_\mathrm{F}v^2eB/\hbar=\Delta_D^2$ (with $l_\mathrm{F}\simeq\mu_0/\hbar\omega_{c1}$) in Eq.\eqref{eq:LLeigen} for $\Delta_D,\Gamma,\gamma\ll \mu_0$ and energies near $\mu_0$; see Appendix \ref{sec:S_effmodel}. Thus Eq.\eqref{eq:LLeigen} reduces to the two doubly degenerate eigenvalues that we use in our analytical calculations
\begin{align}
    \mathcal{E}_{l\pm}&=\left[\epsilon_{1,l}+\epsilon_{2,l}-i\Gamma\pm\sqrt{(\epsilon_{1,l}-\epsilon_{2,l}-i\gamma)^2+\Delta_D^2}\right]/2. \label{eq:LLeigen_Semiclassic}
\end{align}

We note that, in the absence of impurity scattering, our model has a field-induced transition from a gapped insulator
to an gapless metal above a critical field $B_c=\sqrt{m_1m_2}\Delta_D/e\hbar$ [see Appendix \ref{sec:S_Bc}]. 
This can also be seen in Fig.\ref{fig:Model}(b). We focus on the insulating regime $B < B_c$ in this paper.

We list the various symbols related to the different combinations of the parameters of the model, and used in our analysis, for ready
reference in table \ref{tab:Symbols} of Appendix \ref{sec:symbols_S}.

%

\section{Physical picture of $\mathrm{SdH}$ and $\mathrm{dHvA}$ oscillations in insulators}\label{Sec:PhysicalOrigin}

Before turning to the details of our calculations, we present a physical picture to see why the 
the LE-DOS (SdH) and magnetization (dHvA) oscillations in a hybridization-gap insulator differ from each other, and 
why these results are so different from standard quantum oscillations in metals.

First, consider the limit of zero hybridization ($v=0$) in the disorder-free Hamiltonian of Eq.\eqref{eq:Model}, which is
a metal with overlapping electron and  hole bands that cross at $\kf$ at an energy $\mu_0$; see Fig.\ref{fig:schematic}(a,b). 
Both the bands give rise to SdH and dHvA oscillations with same frequency $F_0=(\hbar/2\pi e)\pi\kf^2$ 
corresponding to the area of the semiclassical orbit at $\mu_0=\hbar^2\kf^2/2m_1$. 
LE-DOS oscillations arise due to the $1/B$-periodic passing of LLs across the chemical potential $\mu = \mu_0$.
This occurs whenever $\epsilon_l$ matches $\mu$ and leads to SdH oscillations at frequency $F_0$.
Each time a LL passes through $\mu$, the total number of occupied LLs has a discrete jump leading to 
sharp periodic changes of the total energy $E(B)=N_B\sum_{\epsilon_l\leq \mu}(\epsilon_l-\mu)$, where $N_B=eB/h$ is the LL degeneracy.
As a result, the $T=0$ magnetization $M = - (\partial E/\partial B)$ oscillates as a function of $1/B$ with the same frequency $F_0$. 

Next, consider the LE-DOS oscillations in the hybridization-gap insulator, focusing first on the disorder-free case, with
the chemical potential $\mu_0$ in the gap at the crossing of the unhybridized bands; see Fig.\ref{fig:schematic}(c,d).
The conduction band edge $\mathcal{E}_{c} = \min\mathcal{E}_+(k)$ occurs at $k = k_c$, 
and the valence band edge $\mathcal{E}_{v} = \max\mathcal{E}_-(k)$ at $k = k_v$, with $\mathcal{E}_{c/v}=\mu_0\pm \Delta_I/2$.
The LE-DOS oscillations arise from $1/B$-periodic passage of LLs through the conduction- and valence-band gap edges.
In Fig.\ref{fig:schematic} panel (c) is at a field $B$ and panel (d) at higher field $B'$ corresponding to successive crossing of a LL through 
band edges, i.e., $\mathcal{E}_{l+1,\pm}(B)=\mathcal{E}_{c/v}$ and $\mathcal{E}_{l\pm}(B)=\mathcal{E}_{c/v}$. 
This immediately leads to a $1/B$-periodic modulation of the DOS with frequencies $F_\pm$ determined by $k_c$ and $k_v$ of the gap-edge states,
distinct from $F_0$ corresponding to the unhybridized $k_F$. We show below in Sec.~IV  (and Appendix \ref{sec:S_edgefreq}) the 
SdH oscillations have frequencies $F_\pm=F_0\mp (m_2-m_1)\Delta_I/(4\hbar e)$. Clearly these oscillations need thermal excitation to 
the gap edge, which this leads to an $\exp(- \Delta_I/2T)$ factor in the amplitude. What is less obvious is 
a Dingle factor of $\exp(-\pi\Delta_I T/(\hbar \omega_c)^2)$ that we find in our analysis below.

Impurities lead to in-gap spectral weight~\cite{LF_2018} at $\mu_0$ that leads to oscillations at the unhybridized $F_0$
with a LK-like $T$-dependence. We show below (using a semi-classical saddle point analysis) that the
LE-DOS oscillation is the sum of three pieces, the band-edge oscillations at $F_0\pm \delta F$ and the impurity-induced
oscillations at $F_0$, each with their characteristic $T$-dependence and Dingle factors.

Finally, let us turn to the magnetization oscillations in the disorder-free insulator, which have a very different origin from the LE-DOS oscillations
described above. The total energy  $E(B)=N_B\sum_l (\mathcal{E}_{l-}-\mu)$
is given by a sum over all occupied states $\mathcal{E}_{l-}$ below the chemical potential $\mu$, which is inside the gap.
We next show that there is an unusual aspect  [see Appendix \ref{sec:M_zeroT_S}] to this sum which can be best seen by splitting $\mathcal{E}_{l-}$ into
 $\mathcal{E}_{l-}=\overline{\mathcal{E}}_{l-}+\widetilde{\mathcal{E}}_{l-}$
 with $\overline{\mathcal{E}}_{l-}=(W+\hbar eBl/m_-)/2$ and $\widetilde{\mathcal{E}}_{l-}=-[(W-\hbar eBl/m_+)^2+\Delta_D^2]^{1/2}/2$. 
 This decomposition leads to $E(B) =E_\mathrm{nosc}+E_\mathrm{osc}$. 
It is easy to verify that $E_\mathrm{nosc}=\sum_l[\overline{\mathcal{E}}_{l-}-\mu]$ is a smooth monotonic function of $B$ and the
oscillations arise entirely from  $E_\mathrm{osc}=\sum_l\tilde{\mathcal{E}}_{l-}$.

Thus the dHvA oscillations can be thought to arise from the valence band of a `fictitious' particle-hole symmetric band structure 
$\widetilde{\mathcal{E}}_{\pm}(k)=\pm[(W-\hbar^2k^2/2m_+)^2+\Delta_D^2]^{1/2}/2$. Landau quantization of $\widetilde{\mathcal{E}}_{\pm}(k)$ 
leads to energy levels $\widetilde{\mathcal{E}}_{l\pm}$ for $B\!\neq\!0$ shown in Fig.\ref{fig:schematic}~(e,f).
The total energy $E(B)$ changes abruptly as the energy level $\widetilde{\mathcal{E}}_{l-}$ periodically enters the \emph{electron-like} part of the fictitious valence band from the \emph{hole-like} part through the gap edge (maximum) $\widetilde{\mathcal{E}}_v=-\Delta_D/2$ for some $l$ and $B$. This occurs when
$\widetilde{\mathcal{E}}_{l-}=\widetilde{\mathcal{E}}_{v}$, or equivalently $\hbar eBl/m_+=W$, which leads to
dHvA oscillations with unhybridized frequency $F_0$. 
This frequency corresponds to the semiclassical orbit of area $\pi k_F^2$ originating 
from the gap edge of the fictitious energy dispersion $\tilde{\mathcal{E}}_-(k)$. 
Remarkably, the actual chemical potential $\mu$ plays no role here and 
enters only the non-oscillatory part $E_\mathrm{nosc}$ as long as it lies in the gap.

We note that the same argument also give a simple understanding of the dHvA oscillations in the original KC model~\cite{KC_2015}
where one of the bands has infinite mass . The energy eigenvalues of the KC model can be obtained as the limiting case of Eq.\eqref{eq:LLeigen_Semiclassic} 
for $m_2\to \infty$ and $\Gamma=0$. 

Having obtained physical insight into the origin of quantum oscillations, their frequencies, and the dichotomy between SdH and dHvA oscillations at low temperature in an insulator, we next turn to detailed analytical and numerical calculations that confirm this simple picture [Fig.\ref{fig:schematic}] and extend it to
finite temperature and include Dingle damping.

\section{Low energy DOS} \label{sec:ledos}

In this section we discuss the oscillations in LE-DOS at the chemical potential $\mu_0$, a proxy for SdH oscillations,
which is defined as
\begin{equation}
\label{eq:LEDOS}
D(T)=-\int_{-\infty}^{\infty}d\xi \frac{\partial n_\mathrm{F}(\xi,T)}{\partial \xi} A(\xi).
\end{equation}
The Fermi function $n_\mathrm{F}(\xi,T)=(e^{\beta\xi}+1)^{-1}$ with $\beta\!=\!1/T$  ($k_\mathrm{B}\!=\!1$), and 
the single-particle DOS (per unit area) 
\begin{align}
A(\xi)&=-\left(\frac{N_B}{\pi}\right){\rm Im}\sum_{l,b,p=\pm}\frac{1}{\xi+\mu_0-\mathcal{E}_{l,bp}} \label{eq:DOS}   
\end{align}
is obtained from the complex eigenvalues of Eq.\eqref{eq:LLeigen}, and 
 $N_B=B e/h$ is the LL degeneracy. 

We focus only on the oscillatory part of DOS and LE-DOS, and  
to make analytical progress, we convert the LL sum in Eq.\eqref{eq:DOS} into an integral using the Poisson summation formula. 
In the limit $\mu_0\gg \hbar\omega_{c1}$ using the semiclassical approximation $\mathcal{E}_{l,b\pm}\approx\mathcal{E}_{l\pm}$ [Eq.\eqref{eq:LLeigen_Semiclassic}] 
we obtain
\begin{align}
A(\xi)=&\frac{1}{2\pi^2\hbar^2}\ {\rm Im}\!\sum_{p=\pm ,k\neq 0} \int_{l=0^-}^{\infty}dl \, e^{2\pi\imath kl}\frac{c_p(\xi)}{l-l_p(\xi)}. 
\label{eq:DOS_SemiClassic}
\end{align}
The integer $k$ labels harmonics, and
$l_p(\xi)$ and $c_p(\xi)$ are the poles and residues of $(\xi+\mu_0-\mathcal{E}_{lp})^{-1}$ in the complex $l$-plane [see Appendix \ref{sec:S2}].

For $\mu_0\gg \hbar\omega_{c1}$, we can extend the lower limit of the integral to $-\infty$ since ${\rm Re}(l_\pm)\gg 1$ and the 
poles are far from the origin. The oscillatory part of the DOS is thus
\begin{align}
A(\xi)&=\frac{1}{\pi\hbar^2}{\rm Im}\sum_{k\neq 0,p=\pm}is_p(\xi)c_p(\xi)e^{2\pi\ci ks_p(\xi)l_p(\xi)}, \label{eq:DOS_SemiClassic_2}
\end{align}
with $s_p=\mathrm{sgn}[{\rm Im}(l_p)]$. Substituting this in Eq.\eqref{eq:LEDOS}, we obtain the oscillatory part of $D(T)$ by 
evaluating the energy integral as follows; see  Appendix \ref{sec:S3} for details.

At low temperature $T\ll \Delta_I$, the main contribution comes from two saddle points in the complex $\xi$-plane 
$\widetilde{\xi}_{k\pm}\simeq-\ci\Gamma_c\pm\Delta_I/2+\mathcal{O}(k^2T^2/\hbar^2\omega_{c}^2)$, where 
$\omega_c=eB/(m_1+m_2)$ and $\Gamma_c=(m_1\Gamma_1+m_2\Gamma_2)/(m_1+m_2)$.
In addition, the region near $\xi=0$ on the real axis contributes to the energy integral in Eq.\eqref{eq:LEDOS} when $A(\xi=0)\neq 0$, i.e., 
in the presence of non-zero in-gap DOS for $\Gamma\neq 0$. Incorporating all the contributions, we obtain an expression for 
$D(T)$ by deforming the path of integration from real axis to a suitably chosen contour on the complex plane which passes through 
the two saddle points and the region near $\xi=0$ on the real axis [Appendix \ref{sec:S3}].  

Thus we get $D(T)=D_g(T)+D_0(T)$, where the $D_g(T)$ is gap-edge contribution arising from the two saddle-points and 
$D_0(T)$ is the impurity induced in-gap DOS. The saddle-point contribution 
\begin{align}
D_g(T)&=\frac{1}{\pi \hbar^2}{\rm Re}\sum_{k,p,\zeta}M_{kp\zeta}R_Te^{-\pi k/\omega_c|\tau_{kp}|}e^{\imath 2\pi k s_{kp} (F_\zeta/B)} 
\label{eq:LEDOSg}
\end{align}
corresponds to the oscillations from energy levels passing through the gap edges $\mathcal{E}_{c/v}$ shown in Fig.\ref{fig:schematic}(c,d). 
Here $k$ labels the harmonics, $p = \pm$, $\zeta = \pm$.
The $T$ dependent amplitude $R_T=(\pi \Delta_I/T)^{1/2}\exp{(-\Delta_I/2T)}$ has a Schottky-like activated form controlled by the indirect gap. 
The Dingle damping is controlled by a field, temperature and impurity scattering dependent $1/\vert\tau_{kp}\vert$
where $\hbar/\tau_{kp}= [2\gamma m_+/(m_1+m_2)+pk\pi(\Delta_IT/\hbar\omega_c)]$ with
$\gamma = \Gamma_1 - \Gamma_2$. 
The factor $M_{kp\zeta}$ is given by $M_{kp\zeta}=-\zeta(ps_{kp})^{3/2}\exp{(-\imath\zeta \Gamma_c/T)}(m_1+m_2)/2$
where $s_{kp}=\mathrm{sgn}(\tau_{kp})$.

We emphasize several important features of Eq.~\eqref{eq:LEDOSg}.
The most significant result here is the analytical expression $F_\pm=F_0\mp(m_2-m_1)\Delta_I/(4\hbar e)$
for the oscillation frequencies. How these frequencies originate from the gap-edge states was
discussed in the previous Section (see [Fig.\ref{fig:schematic}(c,d)]. 
In our analysis, they can be traced to the real part of the pole $l_{p}(\tilde{\xi}_{k\zeta})=(F_\zeta/B)+\imath/(2\omega_c\tau_{kp})$
at the complex saddle point. 

\begin{figure}
	\centering
	{
		\includegraphics[width=.45\textwidth]{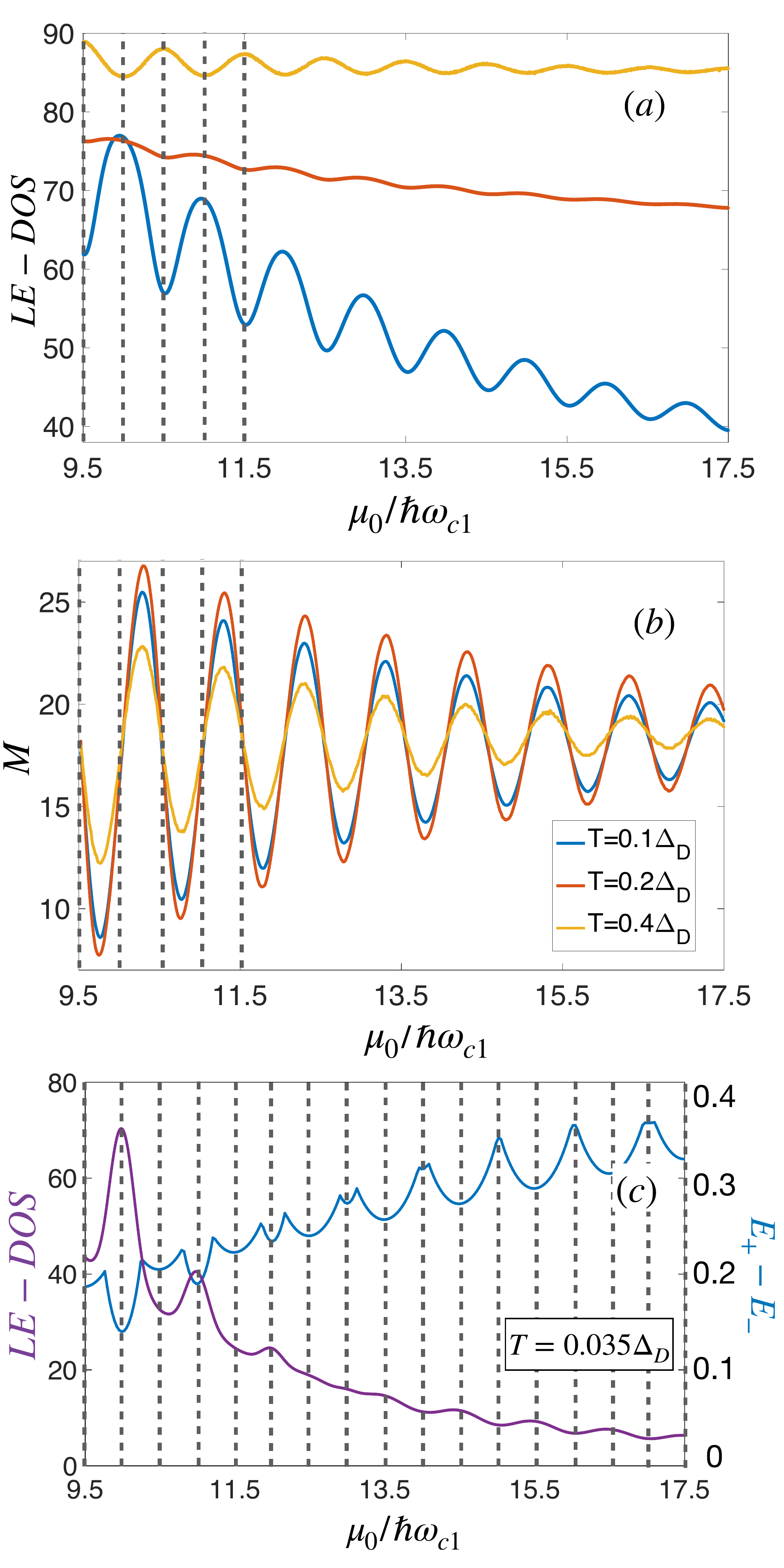}
	}
	\caption{{\bf LE-DOS and magnetization oscillations:} 
   	(a) LE-DOS vs. $1/B$ for $\Gamma=0$ and chemical potential $\mu_0$ at three temperatures indicated in panel (b). The vertical dashed lines are guide to eye for the $\pi$ phase shift between low- and high-temperature oscillations. The oscillation amplitude becomes very small at $T=0.2\Delta_D$, close to $T=T_\pi$ where $\pi$ phase shift occurs.
	(b) Magnetization oscillations for three different temperatures. The amplitude shows non-monotonic temperature dependence. The contrast of dHvA oscillations with LE-DOS oscillations [panel (a)], unlike in a metal, is evident.
	(c) The beat pattern in LE-DOS oscillations at low temperature ($T=0.035\Delta_D$) correlates with the difference $(E_+-E_-)$ in eigen energies closest to the gap edges $\mathcal{E}_{c/v}$ [Fig.\ref{fig:schematic}(e),(f)]. All the results in panels (a), (b) and (c) are obtained using the energy eigenvalues $\mathcal{E}_{l,b\pm}$ in Eq.\eqref{eq:LLeigen}. $E_+-E_-$ in (c) is given in units of $\Delta_D$.
	}
	\label{fig:M_DOS} 
\end{figure}

The two close-by frequencies $F_{\pm}$ give rise to a beat pattern at low $T$. 
We can see this clearly in our numerical results in  {Fig.~\ref{fig:M_DOS}(c)}, which were 
obtained by numerically evaluating $D(T)$ using Eqs.~\eqref{eq:LEDOS} and \eqref{eq:DOS}. 

We analytically show in Appendix \ref{sec:S_edgefreq} that $F_\zeta$'s emerge from the $1/B$-periodic crossing 
of energy levels $\mathcal{E}_{l\pm}$ through the gap edges $\mathcal{E}_{c/v}$ [Fig.~\ref{fig:schematic}(b),(c)]. 
This is also demonstrated in {Fig.~\ref{fig:M_DOS}(c)}, where we plot the difference $(E_+ \!-\!E_-)$ of maximum and minimum energy 
eigenvalues [Eq.~\eqref{eq:LLeigen}] corresponding to the valence and conduction bands as a function of $1/B$. 
The beat pattern in LE-DOS oscillations at low temperature correlates with $(E_+\!-\!E_-)$.

Another important feature of Eq.~\eqref{eq:LEDOSg} is the Dingle damping that arises from the imaginary part of the pole.
Note the unusual $T$ and $B$ dependence of the Dingle factor $\sim\exp{[-k\pi^2(\Delta_IT/\hbar^2\omega_c^2)]}$ in the absence of impurities.
This leads to a Gaussian peak in the Fourier transform (FT) spectrum of the oscillations unlike the usual Lorentzian peak. 

{The low-temperature beat pattern has been alluded to in ref.~\cite{Pal2017} in a different model of hybridization-gap insulator, mostly based on numerical calculations. Here, we give a controlled analytical derivation and clear physical picture [Fig.\ref{fig:schematic}(c)] of the beat frequencies for the first time. Furthermore, we provide the detailed field, temperature and disorder dependence of associated oscillations.}
 
\begin{figure}
	\centering
	{
		\includegraphics[width=0.45\textwidth]{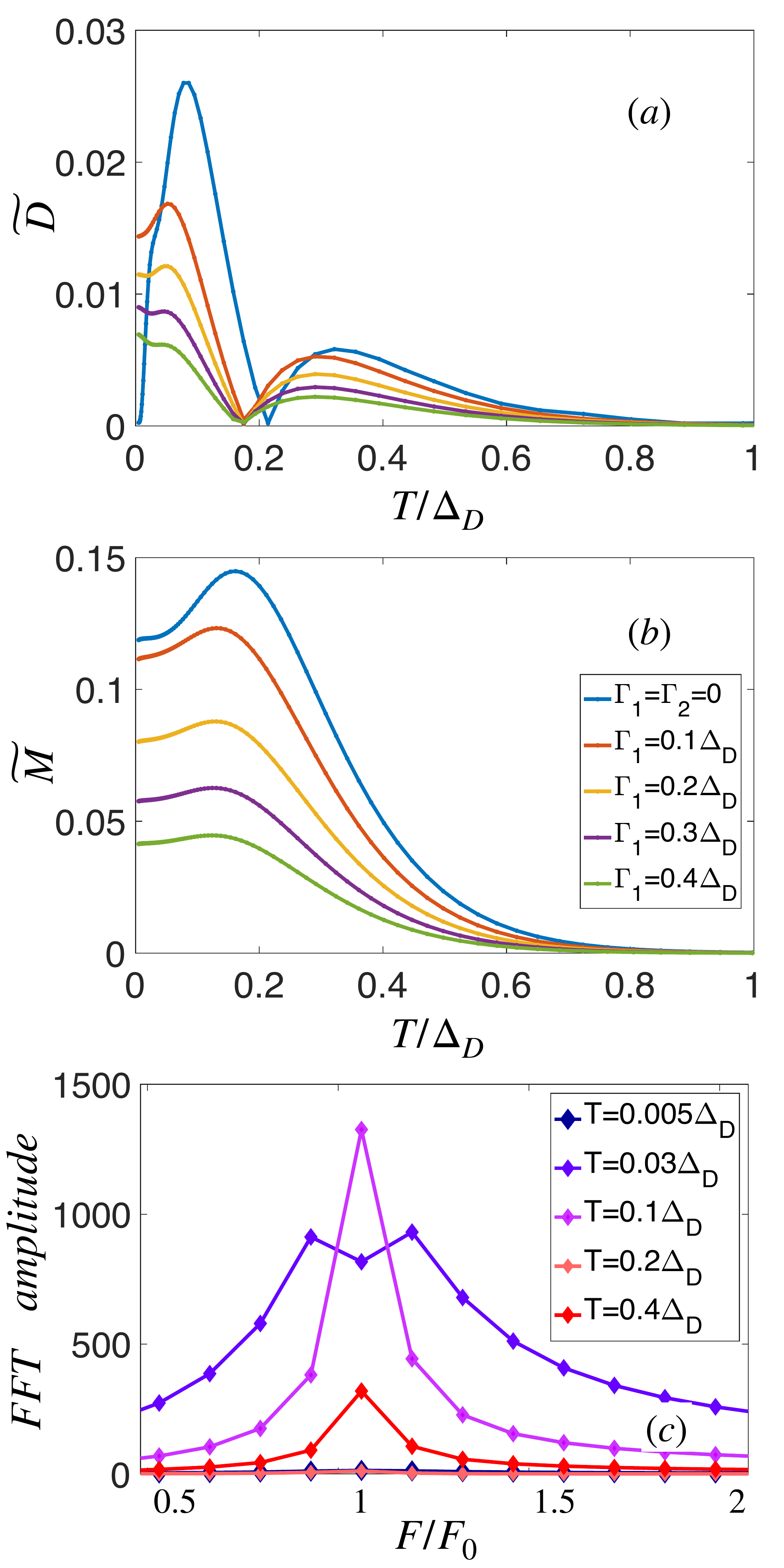}
	}
	
	\caption{{\bf Temperature dependence of LE-DOS and magnetization oscillation amplitude:} (a) LE-DOS oscillation amplitude $\widetilde{D}$ at frequency $F_0$ as a function of temperature for different impurity scattering rates, as indicated in panel (b); $\widetilde{D}$ is extracted from Fourier transform (FT) spectrum of LE-DOS oscillations and is normalized by $T=0$ value of the amplitude $\widetilde{D}_{v=0}(0)$ for zero hybridization. Small scattering rates $\Gamma_1=\Gamma_2=0.0005\Delta_D$ have been used for numerical computation of LE-DOS in the disorder-free case (blue line). For all finite $\Gamma_1$s, $\Gamma_2=0.1\Delta_D$.
	(b) Magnetization oscillation amplitude at $F_0$, $\widetilde{M}(T)$, normalized by its $T=0$ value $\widetilde{M}_{v=0}(0)$ for zero hybridization (see Appendix \ref{sec:M_num_S} for details). 
	The results in panels (a) and (b) are obtained using the energy eigenvalues $\mathcal{E}_{l,b\pm}$ in Eq.\eqref{eq:LLeigen}. (c) FT spectrum at several temperatures for disorder-free system. The peaks at frequencies $F_\zeta$ from gap-edge oscillations is visible at $T=0.03\Delta_D$.
		}
	\label{fig:A_M_T} 
\end{figure}

We next turn to the impurity-induced in-gap LE-DOS, arising from the region near $\xi=0$ in the integral of Eq.\eqref{eq:LEDOS}, which is given by
\begin{align}
D_0(T)=\frac{2}{\pi\hbar^2}\cos\left[2\pi\left(\frac{F_0}{B}\right)\right]\sum_{k,p}\tilde{M}_p\tilde{R}_{T,kp}e^{-\pi k/\omega_c\tilde{\tau}_p} \label{eq:LEDOS0}.
\end{align}
This result is the same as that derived in ref.~\onlinecite{LF_2018}, which however did not obtain $D_g(T)$.
Here $\tilde{M}_p=(1/2)[(m_1+m_2)\Gamma_c/\sqrt{\Gamma_c^2+(\Delta_I/2)^2}+p(m_1-m_2)]$ 
and $\tilde{R}_{T,kp}=\chi/\sinh{\chi}$, with $\chi=2\pi^2\tilde{M}_pkT/\hbar eB$, is an effective LK-like $T$-dependent factor 
governed by both band masses and impurity scattering. The Dingle damping factor is 
$1/\tilde{\tau}_p=[\sqrt{\Gamma_c^2+(\Delta_I/2)^2}+p\Gamma_r]/\hbar$ with $\Gamma_r=(m_1\Gamma_1-m_2\Gamma_2)/(m_1+m_2)$. 

The amplitudes of the LE-DOS oscillations due to gap edges [Eq.\eqref{eq:LEDOSg}] and the in-gap states [Eq.\eqref{eq:LEDOS0}] have completely different temperature dependences. The former is identically zero at $T=0$ and increases in an activated manner with $T$ irrespective of the strength of impurity scattering. In contrast, 
the amplitude of oscillations from in-gap states decreases as a function of $T$ with an effective LK form and is only present for $\Gamma\neq 0$. 

Remarkably, these two contributions coexist as shown by the contour integral calculation above. This analysis, however, is only valid at low temperature $T\ll \Delta_I$. For higher temperatures, $\Delta_I\lesssim T\ll \hbar\omega_{c1}$, we complement our analytical results by 
direct numerical evaluation of Eq.~\eqref{eq:LEDOS}. The results for the LE-DOS as a function of $\mu_0/\hbar\omega_{c1}\propto 1/B$ in the disorder-free 
case are shown in Fig.\ref{fig:M_DOS}(a) for three different temperatures. Similar features are seen for $\Gamma\neq 0$ (not shown). 
We also see, consistent with ref.\onlinecite{FW_2016}, that there is a $\pi$-phase shift of the oscillations at a temperature $T_{\pi}\sim \Delta_I/2$, which 
coincides with the temperature at which the FT amplitude $\widetilde{D}$ vanishes, as shown in Fig.~\ref{fig:A_M_T}(a). 
$T_\pi$ shifts to a slightly lower value for non-zero $\Gamma$. 
The phase shift and vanishing of $\widetilde{D}$ presumably arise from a cancellation between oscillations with different frequencies. 

The presence of the frequencies arising from two gap edges and from the in-gap states can also be seen in our numerical 
FT spectrum in Fig.~\ref{fig:A_M_T}(c). {At higher temperatures $\Delta_I<T<\hbar\omega_{c1}$, the effect of the gap becomes negligible due to thermal excitations and we expect to recover standard oscillations of a metal. Thus, in our numerical results, e.g. the FT spectrum in Fig.~\ref{fig:A_M_T}(c), we see that two frequencies $F_0\pm \delta F$, seen at low temperature, merge into a single frequency $F_0$ at higher temperature.} 

In Fig.~\ref{fig:A_M_T}(a), for the chosen range of values of $\Gamma$, the FT amplitude $\widetilde{D}\equiv\widetilde{D}(F_0)$ at frequency $F_0$ decreases with increasing impurity scattering as expected from the Dingle damping in both $D_g(T)$ [Eq.\eqref{eq:LEDOSg}] and $D_0(T)$ [Eq.\eqref{eq:LEDOS0}]. However, the amplitude $\widetilde{D}(F_0)$ can have much more subtle non-monotonic dependence on both $T$ and $\Gamma$, for different choices of $\Gamma$, as we show in Appendix \ref{sec:nontriv_Amp}. This is because the saddle-point contribution $D_g(T)$ leads to a Gaussian peak at frequency $F_\zeta$ in the FT spectrum and affects the amplitude at the close-by frequency $F_0$ due to its finite width arising from the Dingle damping in Eq.\eqref{eq:LEDOSg}. As a result, LK-like temperature dependence of $D_0(T)$ and activated behaviour of $D_g(T)$ both contribute to temperature dependence of $\widetilde{D}$ in general, leading to complicated non-monotonic $T$ dependence (Appendix \ref{sec:nontriv_Amp}).

\section{Magnetization} \label{sec:magnetization}

In Section III, the dHvA oscillations at $T=0$ were explained in terms of a fictitious particle-hole symmetric gapped spectrum $\tilde{\mathcal{E}}_{\pm}$
[see Fig.~\ref{fig:schematic}(e),(f)] where the semiclassical $k$-space orbits at the gap edges have exactly the same area as the unhybridized crossing 
[Fig.~\ref{fig:schematic}(a),(b)] corresponding to the frequency $F_0$. Here we confirm that the oscillations with frequency $F_0$ persists at 
finite temperature and obtain an analytical expression for the oscillatory part of $M$ for $T\ll \Delta_I,\hbar \omega_c$ via a saddle-point approximation. 
We show that saddle-point for $M$ is completely different from the one that contributes to LE-DOS gap-edge oscillations, and thus 
affirm the unusual dichotomy between dHvA and SdH oscillations in hybridization-gap insulators. 
We corroborate our analytical approximations through numerical calculations which extend to higher temperature.

To compute the magnetization, we use the Matsubara representation of the grand potential ~\cite{KC_2015,Hartnoll_2015} 
$\Omega(T)=-TN_B[\sum_{\omega_n>0,l,bp}\ln{(\mathcal{E}_{l,bp}-\mu_0-\imath \omega_n)}e^{\imath \omega_n0^+}+\mathrm{c.c.}]$
where  $\omega_n=(2n+1)\pi T$ ($n\in\mathbb{Z}$) are fermionic frequencies; see Appendix \ref{subsec:Mamp}. 
In the semiclassical limit $\mu_0\gg \hbar\omega_{c1}$ we can write the 
oscillatory part of magnetization $M(T)=-\partial \Omega(T)/\partial B$ as
\begin{align}
M&=\frac{8\pi T\mu_0}{\hbar \omega_{c1}\phi_0}\sum_{p,k=1}^\infty\left[\sum_{n=0}^\infty F_{kp}(n)\right]. \label{eq:M_Semiclassic}
\end{align}
Here $F_{kp}(n) \equiv F_{kp}(\imath \omega_n)$ is given by the expression
$F_{kp}(n) = \sin{[2\pi k (F_0/B)]} e^{2\pi \imath ks_p(n){\rm Im} [l_p(n)]}$.
$l_p(n)$ denotes the pole $l_p(\xi)$ in Eq.\eqref{eq:DOS_SemiClassic} with $\xi \to \imath \omega_n$ [Sec.\ref{sec:S2} and {SM, Sec.\ref{subsec:Mamp}}] 
and $s_p(n)=\mathrm{sgn}[{\rm Im}\{l_p(n)\}]$.

As shown in {SM, Sec.\ref{subsec:M_Tpeak}}, we evaluate the Matsubara sum in Eq.~\eqref{eq:M_Semiclassic} for $T\ll \Delta_I,\hbar\omega_c$ 
using the Euler-Maclaurin formula
\begin{align}
 T\sum_{n=0}^\infty F_{kp}(n)&\approx \int_0^\infty \frac{d\omega}{2\pi} F_{kp}(\imath \omega)+{T\over 2}\left[F_{kp}(0)+F_{kp}(\infty)\right]
 \label{eq:EulerMacLaurin}  
\end{align}
where we have used $dn=d\omega/(2\pi T)$ and $F_{kp}(n\!\to\!\infty)=0$. 
The integral in the first term does not depend on temperature, and can be evaluated using a saddle-point approximation. 
The saddle point $\widetilde{\omega}=(-\Gamma_c-m_r\Delta_I/2\sqrt{1-m_r^2})$, with $m_r=(m_2-m_1)/(m_1+m_2)$, 
is different from the saddle points that govern the LE-DOS integral [Eq.\eqref{eq:LEDOS}].  
The saddle point here leads to the pole at $l_p(\imath \tilde{\omega})=(F_0/B)+\imath/(2\omega_c\tau_p)$ 
with $1/\tau_{p}= [2m_+\gamma/(m_1+m_2)+(p+m_r^2)\Delta_I/(2\sqrt{1-m_r^2})]/\hbar$. 
The real part of the pole gives rise to an oscillation frequency $F_0$, as if the system has a Fermi surface with an area $\pi\kf^2$ like a metal. 
But, unlike a metal, here the frequency appears from the underlying fictitious particle-hole symmetric gapped system of Fig.~\ref{fig:schematic}(e),(f). 

The temperature dependence in $M(T)$ comes from the next order terms in Eq.\eqref{eq:EulerMacLaurin}. 
Keeping only the leading correction in $T/\sqrt{\hbar \omega_c\Delta_I}$, we obtain 
\begin{align}
M&\propto \sin\left[2\pi k \left(\frac{F_0}{B}\right)\right]\sum_{p,k=1}^\infty \left[\frac{(1-m_r^2)^{3/2}}{\sqrt{k}}e^{-\pi k/\omega_c\vert\tau_p\vert}\right.\nonumber \\
&\left.+\frac{\pi T}{\sqrt{\hbar \omega_c\Delta_I}}e^{-\pi k/\omega_c\tau_{1p}(T)} + \dots\right]. \label{eq:dHvA}
\end{align}
Here we have assumed $\Delta_I>\gamma$ to simplify the expression {[Sec.\ref{subsec:M_Tpeak}, SM]}. This result implies a Dingle damping 
$\exp{(-\pi\Delta_I/2\hbar\omega_c)}$ for the clean system ($\Gamma=0$).  

In Eq.~\eqref{eq:dHvA}, $\tau_{1p}(T)=[(\pi T+\Gamma_c)^2+(\Delta_I/2)^2)^{1/2}+p(\Gamma_r-m_r\pi T)]$ is a temperature dependent damping factor. 
This suggests the existence of a peak in the amplitude of one of the oscillation components ($p\!=\!+$) at a temperature $T_\mathrm{peak}\simeq [m_r\Delta_I/(2\sqrt{1-m_r^2})-\Gamma_c]/\pi$, which shifts towards lower temperature with increasing impurity scattering $\Gamma_c$. The peak eventually goes away when $\Gamma_c\gtrsim \Delta_I$, as one anticipates the impurity-induced DOS to fill up the gap completely in this limit. We note that the low-temperature expansion in Eq.\eqref{eq:dHvA} is not strictly valid at $T\sim T_\mathrm{peak}\sim \Delta_I$, however we expects it to reproduce the qualitative features even at intermediate temperatures. Our numerical results confirms this expectation as we discuss below. The low-$T$ expression  of Eq.\eqref{eq:dHvA} leads to non-LK temperature dependence of dHvA oscillations in the hybridized insulator.

For our numerical calculations in the disorder-free case $\Gamma=0$, we compute $M(T)$ using 
\begin{align}
	\Omega(T)&=-\int_{-\infty}^{\infty}d\xi \frac{\partial n_\mathrm{F}(\xi,T)}{\partial \xi}  \Omega(\xi,T=0), \label{eq:GrandPot_Clean}
\end{align} 
for the grand potential at finite temperature with chemical potential $\mu_0$. Here $\Omega(\xi,T\!=\!0)=N_B\sum'_{lbp}(\mathcal{E}_{l,bp}-\mu_0-\xi)$ is the grand potential or total energy at $T\!=\!0$, where the sum is restricted to $\mathcal{E}_{lbp}\leq \mu_0+\xi$ (see Appendix \ref{sec:M_num_S}). The numerical results for the magnetization oscillations obtained using the energy eigenvalues of Eq.\eqref{eq:LLeigen} with $\Gamma=0$ are shown in Fig.~\ref{fig:M_DOS}(b) as a function of $1/B$ for three temperatures. The oscillations frequency is indeed $F_0$, in agreement with our analytical results in the semiclassical limit. The FT amplitude $\widetilde{M}\equiv \widetilde{M}(F_0)$ at frequency $F_0$ is shown in Fig.~\ref{fig:A_M_T}(b); $\widetilde{M}$ exhibits non-monotonic behaviour with $T$ with a peak at intermediate temperature, as predicted by the low temperature expansion in Eq.\eqref{eq:dHvA}. 

To obtain the magnetization oscillations in disordered system with $\Gamma\neq 0$, we use a semiclassical expression similar to Eq.\eqref{eq:M_Semiclassic}, albeit generalized to incorporate the actual energy eigenvalues [Eq.\eqref{eq:LLeigen}], as discussed in Appendix \ref{sec:M_num_S}. Again, we find $M$ oscillations with unhybridized frequency $F_0$ (not shown). The FT amplitude $\widetilde{M}$ is shown as function of temperature for several $\Gamma_1$ for fixed $\Gamma_2$ in Fig.~\ref{fig:A_M_T}(b). The amplitude shows a peak at intermediate temperature, like $\Gamma=0$ case, however, the peak gets weaker with increasing $\Gamma$, in qualitative agreement with the analytical result [Eq.\eqref{eq:dHvA}].

\section{Discussion and Conclusions}

We have focused in this paper on a minimal model of a hybridization-gap insulator and our results are summarized in the Table~ \ref{tab:Summary}. The physical picture explaining the origin of SdH and dHvA oscillations, and why they differ qualitatively, is summarized in Fig.~\ref{fig:schematic}. In this Section, we conclude with a discussion of the assumptions underlying our model, the universality of our main results, and their possible relation to experiments.

Our results are obtained in an insulating regime when the chemical potential lies in the gap. The insulating nature of the state requires that certain conditions be met. First, we need 
$B < B_c = \sqrt{m_1 m_2}\Delta_D/\hbar e$, the critical field above which the system undergoes an insulator-to-metal transition even in the absence of disorder; 
see Fig.~ \ref{fig:Model} (b). Second, when we include the effects of impurities, we must ensure that they do not drive the system metallic.

The role of impurities in an insulator where a heavy inverted band hybridizes with a light band has been analyzed in detail in ref.~\onlinecite{Skinner2019}. The nature of the impurity bound state wavefunction in such a band structure differs qualitatively from that in ordinary semiconductors and results in a localized ``impurity band''. However, the long-range Coulomb interactions that lead to this behavior are hard to include in the analysis of quantum oscillations. Thus we treat impurity effects following ref.~\cite{LF_2018} as self-energies that arise in an approximation akin to the CPA (coherent potential approximation). 

We focus on the regime of weak disorder broadening $|\Gamma_1 - \Gamma_2| < \Delta_D$, the direct band gap, or else the system enters a semi-metallic regime \cite{LF_2018} as deduced from the real part of the energy eigenvalues of the non-Hermitian Hamiltonian. To check the insulating nature of the weak disorder regime, we have computed the d.c. conductivity at $B=0$ using the Kubo formula within an approximation that includes impurity self-energies in the Greens functions but ignores vertex corrections; see Appendix \ref{sec:Transport}. We find that there is an insulating upturn in the d.c. resistivity ($d\rho/dT < 0$), which nevertheless has a large but finite value at $T=0$ in the disordered system. In the absence of impurities, we would of course get an activated resistivity that diverges at $T=0$.

Our results are based on an insulating gap arising from the hybridization of two bands; though we focused on odd-parity hybridization that is not essential for our analysis.
An important question is the extent to which our results give insight into systems 
where the insulating gap results from interaction as in the Kondo insulators \cite{sebastian_2015,Hartstein2018,Sebastian_2018,Li_Matsuda_2018} or excitonic insulators, which may relevant for the semiconductor superlattices~\cite{RRDu_2019,Samarth_2019}.

We note that, within a mean field theory (MFT) of both these systems, one simply obtains an 
effective two-band model like the one we analyze. The analog of the direct gap $\Delta_D$ in our model is determined by the exciton condensate order parameter in the MFT \cite{Cloizeaux1965,Jerome1967,Halperin1968,Allocca2021} for exciton insulators. Similarly, $\Delta_D$  is determined the hybridization amplitude in the slave-boson MFT of Kondo insulators \cite{HewsonBook1993,ColemanBook2015}. One important difference with our model is that the mean-field order parameters, and thus the resulting hybridization, may have 
non-trivial $B$-dependence, as noted in ref.~\onlinecite{Allocca2021}. However, these authors show that these effects are expected to influence only the higher harmonics of the quantum oscillations and not to modify the characteristic features of fundamental harmonic, which is our main focus.

The experimental situation itself is not very clear at this time, except for the fact quantum oscillations are indeed seen in several different classes of insulators. The dHvA experiments in the Kondo insulator SmB$_6$ exhibit an amplitude that shows ~\cite{sebastian_2015,Hartstein2018} a remarkable increase over the LK form at the lowest temperatures, but such a $T$-dependent amplitude is apparently not seen~\cite{Sebastian_2018,Li_Matsuda_2018} in \ce{YbB_12}.

In semiconductor quantum well experiments the band structure is not ``rigid'', i.e., it changes significantly as the system is gated from a metallic to an insulating regime as a result of the changes in the screening. A model similar to ours should be applicable once the chemical potential lies within the gap. Even in this insulating regime, however, there is an order of magnitude difference in the quantum oscillation frequencies between the two experiments and qualitatively different $T$-dependent amplitudes are seen, LK-like in ref.~\cite{Samarth_2019} but monotonically increasing in $T$ in ref.~\cite{RRDu_2019}. 

Even though none of the existing theories can make quantitative connections with the observed quantum oscillations, we emphasize that any theory of such oscillations in an insulator where the gap results from an effective hybridization will necessarily have to build on the theory of quantum oscillations that is developed here. Our analytical results will serve as a template to incorporate more subtle and exotic effects of interactions, at the very least through frequency-dependent self-energies, in strongly correlated Kondo insulators. The features that we have unearthed through our analytical semiclassical results, and for which we provide a simple physical picture, are universal in so far as the dichotomy between dHvA and SdH oscillation frequencies, the nature of the Dingle damping, the temperature dependence of the amplitudes, as well as the role of disorder in giving an in-gap contribution that adds to the gap-edge oscillations in the low-energy DOS.

\section*{Acknowledgements}
We thank Suchitra Sebastian and Nitin Samarth for useful discussions. SB acknowledges support from SERB (ECR/2018/001742), DST, India and the American Physical Society's International Research Travel Award Program (IRTAP). MR was supported by NSF Materials Research Science and Engineering Center Grant DMR-2011876.

\appendix

\section{Table of symbols}\label{sec:symbols_S}
We list the various symbols used in our paper for ready reference in table \ref{tab:Symbols}.
\begin{widetext}
\begin{center}
\begin{table*}[htb!]
\centering
\begin{tabular}{|p{6cm}|p{6cm}|}
\hline
 Symbol & Expression\\
\hline
Effective masses $m_\pm$ & ${m_1m_2}/({m_2\pm m_1})$ \\
\hline
Mass ratio $m_r$ & ${m_+}/{m_-}=({m_2-m_1})/({m_2+m_1})$ \\
\hline
Unhybridized band crossing energy $\mu_0$ & ${Wm_+}/{m_1}$\\
\hline
Wave vector $k_\mathrm{F}$ corresponding to $\mu_0$ & $(2m_+W)^{1/2}/{\hbar}$\\
\hline
Direct gap  $\Delta_D$ & $2(2m_+W)^{1/2}v/{\hbar}$\\
\hline
Indirect gap  $\Delta_I$ & $\frac{2\sqrt{m_1m_2}}{m_1+m_2}\Delta_D$ \\ \hline 
Impurity scattering rates $\Gamma$, $\gamma$, $\Gamma_c$, $\Gamma_r$ & $\Gamma_1+\Gamma_2,~\Gamma_1-\Gamma_2$, $\frac{m_1\Gamma_1+m_2\Gamma_2}{m_1+m_2}$, $\frac{m_1\Gamma_1-m_2\Gamma_2}{m_1+m_2}$ \\
\hline
Cyclotron frequencies  $\omega_{c1}$, $\omega_{c2}$, $\omega_{c\pm}$, $\omega_c$ &  $\frac{eB}{m_1},$ $\frac{eB}{m_2}$, $\frac{eB}{m_\pm}$, $\frac{eB}{m_1+m_2}$ \\
\hline
\end{tabular}
\caption{Table of symbols}
\label{tab:Symbols}
\end{table*}
\end{center}
\end{widetext}

\section{Effective model in the semi-classical limit}\label{sec:S_effmodel}
   In the energy eigenvalues $\mathcal{E}_{lb\pm}$ [Eq.\eqref{eq:LLeigen}], the hybridization term $8lv^2eB/\hbar$ becomes important for $l\approx l_\mathrm{F}$ corresponding to the unhybridised band crossing, i.e. $\epsilon_{1,l_\mathrm{F}}\approx\epsilon_{2,l_\mathrm{F}}\approx\mu_0$. LL energies for $l$ farther from this energy tends to the original unhybridised energies $\epsilon_{1,l}-\imath\Gamma_1$ and $\epsilon_{2,l}-\imath\Gamma_2$. As a result, in the semiclassical limit $\mu_0=Wm_+/m_1\gg \hbar\omega_{c1}$, $l_\mathrm{F}\simeq Wm_+/(\hbar eB)\gg 1$, and we get
   \begin{align*}
  8lv^2eB/\hbar \approx 8l_\mathrm{F}v^2eB/\hbar\simeq\frac{8v^2m_+}{\hbar^2}.W=\Delta_D^2.
   \end{align*}
   The above leads to the \emph{semiclassical} energy eigenvalues of Eq.\eqref{eq:LLeigen_Semiclassic}.
   
\section{Critical field $B_c$ for field-induced insulator to metal transition}\label{sec:S_Bc}
Here we give an estimate \cite{FW_2016} of the critical field $B_c$ for $\Gamma=0$. $B_c$ is obtained from the field at which the minimum, $E_+(B)$, and maximum, $E_-(B)$, of the energy eigenvalues $\mathcal{E}_{l,\downarrow\uparrow +}$ and $\mathcal{E}_{l,\uparrow\downarrow-}$, marked respectively in blue and yellow in Fig.\ref{fig:Model}(b), coincide. We obtain $E_\pm$ from $\partial\mathcal{E}_{l,\downarrow\uparrow +}/\partial l=\partial\mathcal{E}_{l,\uparrow\downarrow -}/\partial l=0$. To this end, for example, we rewrite 
\begin{align}
\label{eq:s2}
\mathcal{E}_{l\uparrow\downarrow -}& = \mu_0+\frac{1}{2}\left[\frac{m_+}{m_-}y-\sqrt{y^2+ay+b} \right.\nonumber \\
&\left.+\frac{\hbar\omega_{c+}}{2}\left(1-\frac{m_+^2}{m_-^2}\right)\right]
\end{align}
from Eq.\eqref{eq:LLeigen}, using $y=\left(\hbar\omega_{c+}l+\hbar\omega_{c-}/2-W\right)$, $a=8v^2eB/\hbar^2\omega_{c+}$, and $b=(8v^2eB/\hbar^2 W\omega_{c+} - 4v^2eB/\hbar m_+ m_-)$, where $\hbar\omega_{c\pm}=eB\hbar/m_\pm$. 
Now, minimizing the above with respect to $y$ or $l$, we obtain
\begin{equation}
\label{70}
y=-a/2+\frac{m_+}{m_-}\sqrt{\frac{b-(a/2)^2}{1-\frac{m_+^2}{m_-^2}}}
\end{equation}
for the maximum of $\mathcal{E}_{l\uparrow\downarrow -}$ in the weak hybridization limit $W\gg m_+v^2/\hbar$,

\begin{align}
&E_-(B)\simeq\mu_0+1/2\left[\frac{m_+}{m_-}\left(\frac{-4v^2eB/\hbar}{\hbar\omega_{c+}}\right)+ \frac{\hbar\omega_{c+}}{2}\left(1-\frac{m_+^2}{m_-^2}\right)\right.\nonumber \\
&\left.-\left(1-\frac{m_+^2}{m_-^2}\right)^{1/2}\left(\frac{8Wv^2eB/\hbar}{\hbar\omega_{c+}}\right)^{1/2}\right]
\end{align}
Following similar steps, the minimum of $\mathcal{E}_{l,\downarrow\uparrow +}$ is obtained as
\begin{align}
&E_+(B)\simeq\mu_0+1/2\left[\frac{m_+}{m_-}\left(\frac{-4v^2eB/\hbar}{\hbar\omega_{c+}}\right)+ \frac{\hbar\omega_{c+}}{2}\left(1-\frac{m_+^2}{m_-^2}\right)\right.\nonumber \\
&\left.+\left(1-\frac{m_+^2}{m_-^2}\right)^{1/2}\left(\frac{8Wv^2eB/\hbar}{\hbar\omega_{c+}}\right)^{1/2}\right]
\end{align}
Using the condition $E_+(B_c)=E_-(B_c)$, we obtain the critical field
\begin{equation}
B_c=\frac{2\Delta_{D}}{e\hbar} \frac{m_+m_-}{\sqrt{m_-^2-m_+^2}}.
\end{equation}
   
   \section{Frequency of DOS oscillations at the gap edges}\label{sec:S_edgefreq}
 We show that the energy levels for non-zero magnetic field periodically crosses through the hybridization gap edges $\mathcal{E}_c$ and $\mathcal{E}_v$ [Fig.\ref{fig:schematic}(c),(d)], i.e. the minimum of the conduction band $\mathcal{E}_+(k)$ and the maximum of valence band $\mathcal{E}_-(k)$, as a function of $1/B$. In the semiclassical limit $\mu_0\gg \hbar\omega_{c1}$, we estimate $\mathcal{E}_{c/v}$ from energy dispersion
\begin{align}
\mathcal{E}_\pm(k)=\frac{1}{2}\left[W+\frac{\hbar^2k^2}{2m_-}\pm\sqrt{\left(\frac{\hbar^2k^2}{2m_+}-W\right)^2+\Delta_D^2}\right] ,
\end{align}
which corresponds to the semiclassical eigenvalues $\mathcal{E}_{l\pm}$ in Eq.\eqref{eq:LLeigen_Semiclassic}. The wavevectors (magnitude) $k_c$ and $k_v$ at the energies $\mathcal{E}_{c/v}$ are obtained from $[\partial \mathcal{E}_{+}(k)/\partial k]_{k=k_c}=[\partial \mathcal{E}_{-}(k)/\partial k]_{k=k_v}=0$ as $ k_{c/v}^2=\kf^2\mp \Delta_I(m_2-m_1)/(2\hbar^2)$. These lead to
\begin{align}
    \mathcal{E}_{c/v}=\mu_0\pm \frac{\Delta_I}{2}
\end{align}
Equating the above with energy levels $\mathcal{E}_{l\pm}$ for two successive LL index, e.g. $l$ and $l+1$, at two fields $B'$ and $B$ ($B<B'$), i.e. $\mathcal{E}_{l,\pm}(B')=\mathcal{E}_{c/v}$ and $\mathcal{E}_{l+1,\pm}(B)=\mathcal{E}_{c/v}$, we obtain
\begin{align}
\frac{1}{F_\pm}&\equiv\frac{1}{B}-\frac{1}{B'}=\frac{\hbar e}{\left(W\mp\frac{\Delta_D^2}{\sqrt{m_-^2/m_+^2-1}}\right)m_+}\nonumber \\
&=\frac{1}{F_0\mp\Delta_I(m_2-m_1)/(4\hbar e)}
\end{align}
The above proves the $1/B$-periodicity of the gap-edge crossing of the energy levels with frequencies $F_\pm$ even though the eigenvalues $\mathcal{E}_{l\pm}$ do not have canonical equispaced LL form. The frequencies arise from the semiclassical orbits of areas $\pi k_c^2$ and $\pi k_v^2$ at energies $\mathcal{E}_{c/v}$ [Fig.\ref{fig:schematic}(c,d)]. The DOS oscillates with $1/B$ periodicity at the gap edges, which are the lowest energy excitations $\pm \Delta_I/2$ away from the chemical potential $\mu_0$. Thus the gap edges contribute to the oscillations of the LE-DOS of Eq.\eqref{eq:LEDOS} with a thermally activated amplitude $\sim \exp{(-\Delta_I/2T)}$ at low temperature in agreement with the low-$T$ saddle-point expression [Eq.\eqref{eq:LEDOSg}]. 

\section{DOS in the semiclassical limit}\label{sec:S2}
Using the eigen energies in the semi-classical limit from Eq.\ref{eq:LLeigen_Semiclassic}, the DOS can be written as 
\begin{align}
A(\xi)&=-2\left(\frac{N_B}{\pi}\right)\mathrm{Im}\sum_{l,p=\pm}\frac{1}{\xi+\mu_0-\mathcal{E}_{l,p}} \label{eq:DOS1}   
\end{align}
The factor of two is due to the degeneracy of the energy levels. Using the expressions for $\mathcal{E}_{l\pm}$ we can write
   \begin{align*}
   &\sum_{p=\pm 1}\left(\frac{1}{\xi+\mu_0-\mathcal{E}_{lp}}  \right)\\
   &=\sum_{p=\pm 1}\left(\frac{1}{\xi+\mu_0-\frac{1}{2}\left[ \epsilon_{1,l}'+\epsilon_{2,l}'+p \sqrt{\left(\epsilon_{1,l}'-\epsilon_{2,l}' \right)^2+\Delta_D^2} \right]} \right)\\
   &=\frac{b_1l+b_0}{a_2l^2+a_1l+a_0}
   \end{align*}
 
   where, $a_2=-\hbar\omega_{c1} \hbar\omega_{c2}$, $a_1=\hbar\omega_{c1}(W-\imath\Gamma_2) +\hbar\omega_{c2}\imath\Gamma_1-(\xi+\mu_0)\hbar\omega_{c-}$, $a_0=-\imath\Gamma_1(W-\imath\Gamma_2)-(\xi+\mu_0)(W-\imath\Gamma_1-\imath\Gamma_2)+(\xi+\mu_0)^2-\Delta_D^2/4$, $b_1=-\hbar\omega_{c-}$ and $b_0=2(\xi+\mu_0)-W+\imath(\Gamma_1+\Gamma_2)$.
We can rewrite the above equation as
   \begin{align}
   \label{eq:s10}
   \sum_{p=\pm }\left(\frac{1}{\xi+\mu_0-\mathcal{E}_{lp}}  \right)
   &=-\frac{1}{2eB\hbar}\sum_{p=\pm}\frac{c_p}{l-l_p}
   \end{align}
   with $l_{\pm}(\xi)=(-a_1\mp\sqrt{a_1^2-4a_2a_0})/2a_2$ and $c_\pm=(-eB\hbar/a_2)[b_1\mp\{b_1(l_++l_-)+2b_0\}/(l_--l_+)]$. Since $\mu_0\hbar\omega_{c+}=W\hbar\omega_{c1}$ and $\hbar^2k_F^2/2m_1=\mu_0$, we get
   \begin{widetext}
   \begin{align*}
  \frac{-a_1}{2a_2}=&\frac{\hbar\omega_{c1}(W-\imath\Gamma_2) +\hbar\omega_{c2}\imath\Gamma_1-\left(\xi+\mu_0\right)\hbar\omega_{c-}}{2\hbar\omega_{c1}\hbar\omega_{c2}}
   =\frac{\hbar^2k_F^2+\imath(m_1\Gamma_1-m_2\Gamma_2)+\left(m_1-m_2\right)\xi}{2eB\hbar}
   \\a_1^2-4a_2a_0=&\left(\hbar\omega_{c1}(W-\imath\Gamma_2) +\hbar\omega_{c2}\imath\Gamma_1-\left(\xi+\mu_0\right)\hbar\omega_{c-}\right)^2\\&-4\hbar\omega_{c1} \hbar\omega_{c2} \left(\imath\Gamma_1(W-\imath\Gamma_2)+\left(\xi+\mu_0\right)(W-\imath\Gamma_1-\imath\Gamma_2)-\left(\xi+\mu_0\right)^2+\Delta_D^2/4\right)
   \\=&\left[ \left(\xi+\mu_0\right)\hbar\omega_{c+}+\imath\Gamma_1\hbar\omega_{c2}+ \imath\Gamma_2\hbar\omega_{c1} -W\hbar\omega_{c1} \right]^2
   -\hbar\omega_{c1} \hbar\omega_{c2}\Delta_D^2
   \\=&\left[ \xi\hbar\omega_{c+}+\imath\Gamma_1\hbar\omega_{c2}+ \imath\Gamma_2\hbar\omega_{c1}  \right]^2
   -\hbar\omega_{c1} \hbar\omega_{c2}\Delta_D^2
   \\\frac{\sqrt{a_1^2-4a_2a_0}}{2a_2}=&-\frac{\sqrt{\left[ (m_1+m_2)\xi+\imath(m_1\Gamma_1+m_2\Gamma_2)\right]^2-m_1m_2\Delta_D^2}}{2eB\hbar}
   \end{align*}
  
    Using the above, we obtain the expression for $l_p(\xi)$ ($p=\pm\equiv \pm 1$) as
\begin{subequations}
   \begin{align}
   l_p(\xi)&=\frac{F_0}{B}+x_p(\xi) \label{eq:poles}\\
   x_p(\xi)&=\frac{\imath(m_1\Gamma_1-m_2\Gamma_2)+\left(m_1-m_2\right)\xi+p \sqrt{\left[ (m_1+m_2)\xi+ \imath(m_1\Gamma_1+m_2\Gamma_2)\right]^2-m_1m_2\Delta_D^2}}{2eB\hbar} \nonumber \\
   &=\frac{1}{2\hbar\omega_{c}}\left[-m_{r}\xi+i\Gamma_{r}\pm\sqrt{(\xi+i\Gamma_c)^{2}-(\Delta_I/2)^2}\right].\label{eq:xpm}
   \end{align}
 Similarly, $c_p(\xi)$ is given by
 \begin{align}
 c_p(\xi)&=(m_1-m_2)+p(m_1+m_2)\frac{\left[\left(m_1+m_2\right)\xi+\imath(m_1\Gamma_1+m_2\Gamma_2)\right]}{\sqrt{\left[ (m_1+m_2)\xi+ \imath(m_1\Gamma_1+m_2\Gamma_2)\right]^2-m_1m_2\Delta_D^2}} \nonumber \\
 &=(m_1+m_2) \left(-m_r+p\frac{\xi+\imath\Gamma_c}{\sqrt{\left[ \xi+ \imath\Gamma_c\right]^2-(\Delta_I/2)^2}} \right) \label{eq:cpS}
 \end{align}
\end{subequations}
 \end{widetext}
Here $\omega_c=eB/(m_1+m_2)$, $m_r=(m_2-m_1)/(m_2+m_1)$, $\Gamma_r=(m_1\Gamma_1-m_2\Gamma_2)/(m_1+m_2)$ and $\Gamma_c=(m_1\Gamma_1+m_2\Gamma_2)/(m_1+m_2)$.
Converting the LL sum over $l$ in Eq.\ref{eq:DOS1} into an integral using Poisson summation formula and evaluating the integrals using the poles $l_p(\xi)$ and the residues $c_p(\xi)$, we obtain the oscillatory part of the DOS as
 \begin{align}
 A(\xi)=&\frac{1}{\pi\hbar^2}\mathrm{Im} \sum_{k\neq 0,p=\pm}\imath s_p(\xi)c_p(\xi)e^{2\pi\imath ks_p(\xi)l_p(\xi)}\label{eq:DOS_semiclassicS}
\end{align}
 with $s_p(\xi)=\mathrm{sgn}[\mathrm{Im}\{l_p(\xi)\}]$.

\section{LE-DOS oscillations at low temperatures}\label{sec:S3}
\begin{figure*}[htb!]
	\centering
	{
		\includegraphics[width=0.7\textwidth]{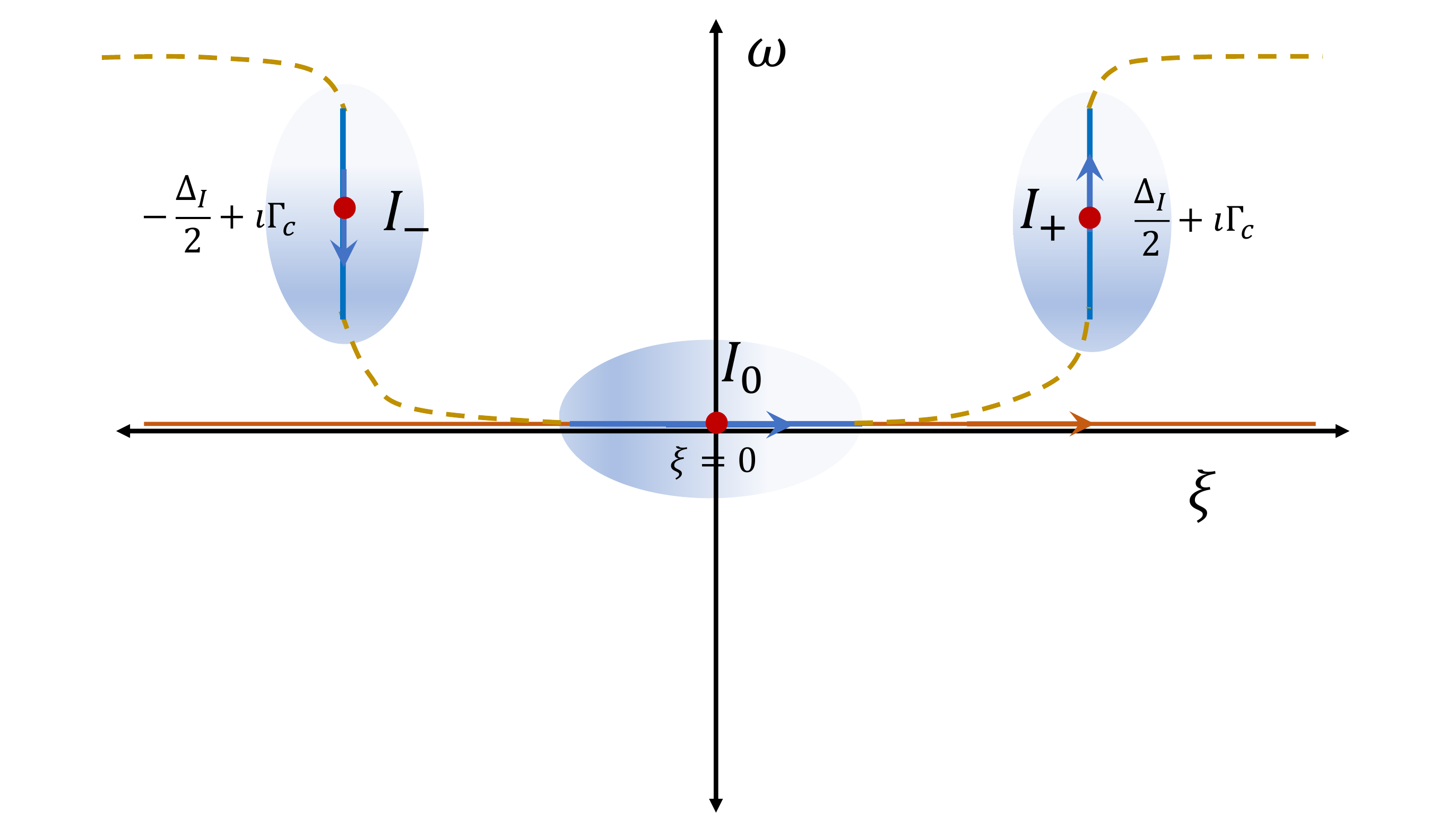}
	}
	
	\caption{{\bf Integration contour for the saddle-point approximation to LE-DOS:} The original integration path (horizontal solid orange line) for LE-DOS energy integral is along the real axis $\xi$ on the complex plane $z=\xi+\imath \omega$. The contour is deformed to go through the saddle points $\tilde{\xi}_{k\pm}\simeq \pm (\Delta_I/2)+\imath \Gamma_c$ and the origin $\xi=0,\omega=0$. Only the integrals $I_\pm$ along the saddle-point paths (vertical solid blue lines) close to the saddle points matter at low temperature $T\ll \Delta_I$. In addition, the integral $I_0$ from the region close to the origin contributes to LE-DOS for finite impurity scattering $\Gamma\neq 0$. The saddle-point paths are vertical for $\Gamma\to 0$, as shown here. The three integration regions $I_\pm,I_0$ are connected by arbitrary contours (dashed yellow lines) which have negligible contribution to LE-DOS at low $T$.
	}
	\label{fig:ContourS} 
\end{figure*}


 

In the semiclassical limit, we rewrite the LE-DOS of Eq.\eqref{eq:LEDOS} using the DOS of Eq.\eqref{eq:DOS_semiclassicS} as
\begin{subequations}
\begin{align}
&D(T)=\frac{1}{2T}\int_{-\infty}^\infty d\xi\frac{1}{1+\cosh\left(\frac{\xi}{T}\right)}A(\xi)\nonumber\\
&=\mathrm{Re}\left[\int_{-\infty}^{\infty}\frac{d\xi}{4T\pi \hbar^2}\sum_{k\neq 0,p}e^{2\pi iks_{p}\left(\frac{F_0}{B}\right)}s_{p}(\xi)c_{p}(\xi)e^{-f_{kp}(\xi)}\right]\label{eq:LEDOS_S},
\end{align}
where
\begin{align}
f_{kp}(\xi)= &- 2\pi \imath ks_p(\xi)x_p(\xi)+2\ln\left[\cosh(\xi/2T)\right], 
\end{align}
\end{subequations}
and we have used $2\cosh^2(\xi/2T)=1+\cosh(\xi/T)$.
We split the integral in Eq.\eqref{eq:LEDOS_S} into three parts [Fig.\ref{fig:ContourS}] as 
\begin{align}
\int_{-\infty}^\infty d\xi [\dots]&=\left(\int_{-\infty}^{-\xi_c}+\int_{\xi_c}^\infty +\int_{-\xi_c}^{\xi_c}\right)d\xi[\dots]\nonumber\\
&=I_++I_-+I_0, \label{eq:IntegralSplitS}
\end{align}
where $\xi_c\gtrsim \Delta_I/2$ is an arbitrary cutoff. We show below that the integrals $I_\zeta$ ($\zeta=\pm$) can be well approximated via a saddle-point method, whereas $I_0$ gets the main contribution from the region near $\xi=0$ at low temperatures $T\ll \Delta_I$. The saddle-point contribution to the LE-DOS [Eq.\eqref{eq:LEDOSg}] is $D_g(T)=I_++I_-$ and impurity induced in-gap DOS [Eq.\eqref{eq:LEDOS0}] $D_0(T)=I_0$.

\subsection{LE-DOS oscillations from the gap edges} \label{Sec:DgS}
To carry out the saddle-point integration for $I_\zeta$, we expand $2\ln[\cosh(\xi/2T)]\simeq \zeta \xi/T-2\ln 2+2\exp{(-\zeta \xi/T)}+\dots$ at low temperature, where $\zeta=1$ ($+$) for $\xi>0$ and $\zeta=-1$ ($-$) for $\xi<0$. Thus, approximating 
\begin{align}\label{eq:fpS}
f_{kp}(\xi)\approx -2\pi k\imath s_p(\xi)x_p(\xi)+\zeta \xi/T+2\ln 2,
\end{align}
 the saddle point is obtained from $\partial f_p(\xi)/\partial \xi=0$ assuming that the sign $s_p(\xi)$ does not vary around the saddle points. For, $T\ll \hbar \omega_c$, we obtain two saddle points for each $k,p$
\begin{align}
\tilde{\xi}_{k\zeta}\simeq-\imath \Gamma_c+\zeta \frac{\Delta_I}{2}\left(1-\frac{k^2\pi^2T^2}{2\hbar^2\omega_c^2}\right) \label{eq:SaddlePointS}
\end{align}
which are complex for $\Gamma\neq 0$. The saddle-point leads to
\begin{subequations} \label{eq:xpcpS}
\begin{align}
x_p(\tilde{\xi}_{k\zeta})&\simeq-\zeta\frac{(m_2-m_1)\Delta_I}{4\hbar e B}+\frac{\imath}{2\omega_c\tau_{kp}}\\
c_p(\tilde{\xi}_{k\zeta})&\simeq -\imath p\zeta \frac{\hbar e B}{\pi kT}
\end{align}
with
\begin{align}
\frac{1}{\tau_{kp}}&=\frac{1}{\hbar}\left[\frac{2\gamma m_+}{m_1+m_2}+pk\pi\frac{\Delta_IT}{\hbar\omega_c}\right],
\end{align}
\end{subequations}
where $s_p(\tilde{\xi}_{k\zeta})=\mathrm{sgn}(\tau_{kp})$. The above implies that the saddle-point value of the pole
\begin{align}
l_p(\tilde{\xi}_{k\zeta})&=\frac{F_\zeta}{B}+\frac{\imath}{2\omega_c\tau_{kp}}
\end{align}
in the DOS [Eq.\eqref{eq:DOS_semiclassicS}] dominates the integrals $I_\zeta$ in the LE-DOS [Eq.\eqref{eq:LEDOS_S}] at low temperature. The real part of the saddle-point pole modifies the frequency of oscillations to $F_\zeta=F_0-\zeta(m_2-m_1)\Delta_I/(4\hbar e)$. 
As shown in Fig.\ref{fig:ContourS}, to evaluate the integrals $I_\zeta$ using the saddle-points in Eq.\eqref{eq:SaddlePointS}, we deform the integration contour from the real axis to the complex plane $z=\xi+\imath \omega$ such that it goes through the saddle points. The deformed path is chosen such that, close to $\tilde{\xi}_{k\zeta}$, the imaginary part of $f_{kp}(z)$ remains constant and the real part has a maximum at the saddle point along the path. This is achieved by the expansion $f_{kp}(z=\xi_{k\zeta}+\eta)\simeq f_{kp}(\tilde{\xi}_{k\zeta})+(1/2)(\partial^2f_{kp}/\partial z^2)_{z=\tilde{\xi}_{kp}}\eta^2$, where $(\partial^2f_{kp}/\partial z^2)_{z=\tilde{\xi}_{kp}}=2ps_{kp\zeta}(\hbar \omega_c)^2/(\pi^2k^2T^3\Delta_I)$ with $s_{kp\zeta}=s_p(\tilde{\xi}_{k\zeta})$, such that
\begin{widetext}
\begin{align}
D_g(T)&=\frac{1}{4T\pi\hbar^2}\mathrm{Re}\left[\sum_{kp\zeta}e^{2\pi\imath k s_{kp\zeta}\left(\frac{F_0}{B}\right)}s_{kp\zeta}c_p(\tilde{\xi}_{k\zeta})e^{-f_{kp}(\tilde{\xi}_{k\zeta})} \int d\eta e^{-\frac{1}{2}\left(\frac{\partial^2f_{kp}}{\partial z^2}\right)_{\tilde{\xi}_{kp}}\eta^2}\right].\label{eq:LEDOSgS}
\end{align}
This leads to the Gaussian integral
\begin{align*}
\int d\eta e^{p s_{kp\zeta}[(\hbar \omega_c)^2/(\pi^2k^2T^3\Delta_I)]\eta^2}&\approx (-ps_{kp\zeta})^{1/2}\int_{-\infty}^\infty d\tau e^{-[(\hbar \omega_c)^2/(\pi^2k^2T^3\Delta_I)]\tau^2} =(-ps_{kp\zeta})^{1/2}\pi^{3/2} k\frac{T^{3/2}\Delta_I^{1/2}}{\hbar \omega_c},
\end{align*}
\end{widetext}
where the integration contour through the saddle-point is chosen via the variable transformation $ps_{kp\zeta}\eta^2=-\tau^2$. For $\Gamma\to 0$, $s_{kp\zeta}=p$ and the saddle-point paths are vertical [Fig.\ref{fig:ContourS}]. Finally, using the above and Eqs.\eqref{eq:fpS},\eqref{eq:xpcpS} in Eq.\eqref{eq:LEDOSgS} we obtain the expression for $D_g(T)$ [Eq.\eqref{eq:LEDOSg}].

\subsection{LE-DOS oscillations from impurity-induced in-gap DOS} \label{Sec:D0S}
The gap-edge oscillations coexist with the in-gap DOS oscillations in the presence of disorder ($\Gamma\neq 0$), and they can be separated from each other at low temperature since splitting of LE-DOS integral in Eq.\eqref{eq:IntegralSplitS} into three independent integrals is well controlled for $T\ll \xi_c\sim \Delta_I$. The in-gap DOS oscillations were derived in ref.\onlinecite{LF_2018}. Here we briefly sketch the derivation for the sake of completeness.

The main effect of impurity-induced DOS arise near $\xi=0$ at the chemical potential $\mu_0$ inside the gap. This is captured by the integral $I_0=D_0(T)=\int_{-\xi_c}^{\xi_c}d\xi[\dots]$ in the LE-DOS integral [Eq.\eqref{eq:IntegralSplitS}]. At low temperatures, for $\Gamma\neq 0$, due to the $[1+\cosh(\xi/T)]^{-1}$ term in Eq.\eqref{eq:LEDOS_S}, the main contribution to $I_0$ comes from the region near $\xi=0$ along the real axis [Fig.\ref{fig:ContourS}]. Thus, by expanding $c_p(\xi)\simeq c_p(0)+c'_p(0)\xi$, $x_p(\xi)=x_p(0)+x'_p(0)\xi$, we can approximate $D_0(T)$ as
\begin{widetext}
\begin{align}
D_0(T)&\simeq\frac{1}{2T\pi\hbar^2}\mathrm{Re}\left[\sum_{kp}s_p(0)e^{2\pi \imath k s_p(0)\left(\frac{F_0}{B}\right)}\int_{-\infty}^\infty d\xi \frac{c_p(0)}{1+\cosh{\left(\frac{\xi}{T}\right)}}e^{2\pi\imath k s_p(0)[x_p(0)+x'_p(0)\xi]}\right]
\end{align}
\end{widetext}
To evaluate the integral above we use the identity~\cite{LF_2018},
\begin{align}
\int_{-\infty}^\infty d\xi \frac{1}{1+\cosh{\left(\frac{\xi}{T}\right)}}e^{2\pi\imath k s_p(0)x'_p(0)\xi}=\frac{4\pi^2 kT^2x_p'(0)}{\sinh(2\pi^2 kT x_p'(0))}. \nonumber
\end{align}
Moreover, from Eqs.\eqref{eq:xpm},\eqref{eq:cpS}
\begin{align*}
x_p(0)&=\frac{\imath}{2\hbar \omega_c} \left[\Gamma_r+p\sqrt{\Gamma_c^2+(\Delta_I/2)^2}\right]\\ x_p'(0)&=\frac{1}{2\hbar\omega_c}\left[-m_r+\frac{p\Gamma_c}{\sqrt{\Gamma_c^2+(\Delta_I/2)^2}}\right]\\  
c_p(0)&=\left[\frac{p(m_1+m_2)\Gamma_c}{\sqrt{\Gamma_c^2+(\Delta_I/2)^2}}+(m_1-m_2)\right].
\end{align*}
 Using the above we obtain the expression for $D_0(T)$ given in Eq.\eqref{eq:LEDOS0}. Here $s_p(0)=\mathrm{Im}[x_p(0)]=p$ since $\sqrt{\Gamma_c^2+(\Delta_I/2)^2}>\Gamma_r$. Moreover, it can be shown that next order in temperature correction appears at $\mathcal{O}(T/\Delta_I,T/\Gamma)$ to $D_0(T)$. $D_0(T)\to 0$ as $\Gamma\to 0$, i.e. for the disorder-free case, as can be verified from Eq.\eqref{eq:LEDOS0}.

\section{Non-trivial temperature dependence of LE-DOS amplitude}\label{sec:nontriv_Amp}
\begin{figure}[htb!]
	\centering
	{
		\includegraphics[width=0.4\textwidth]{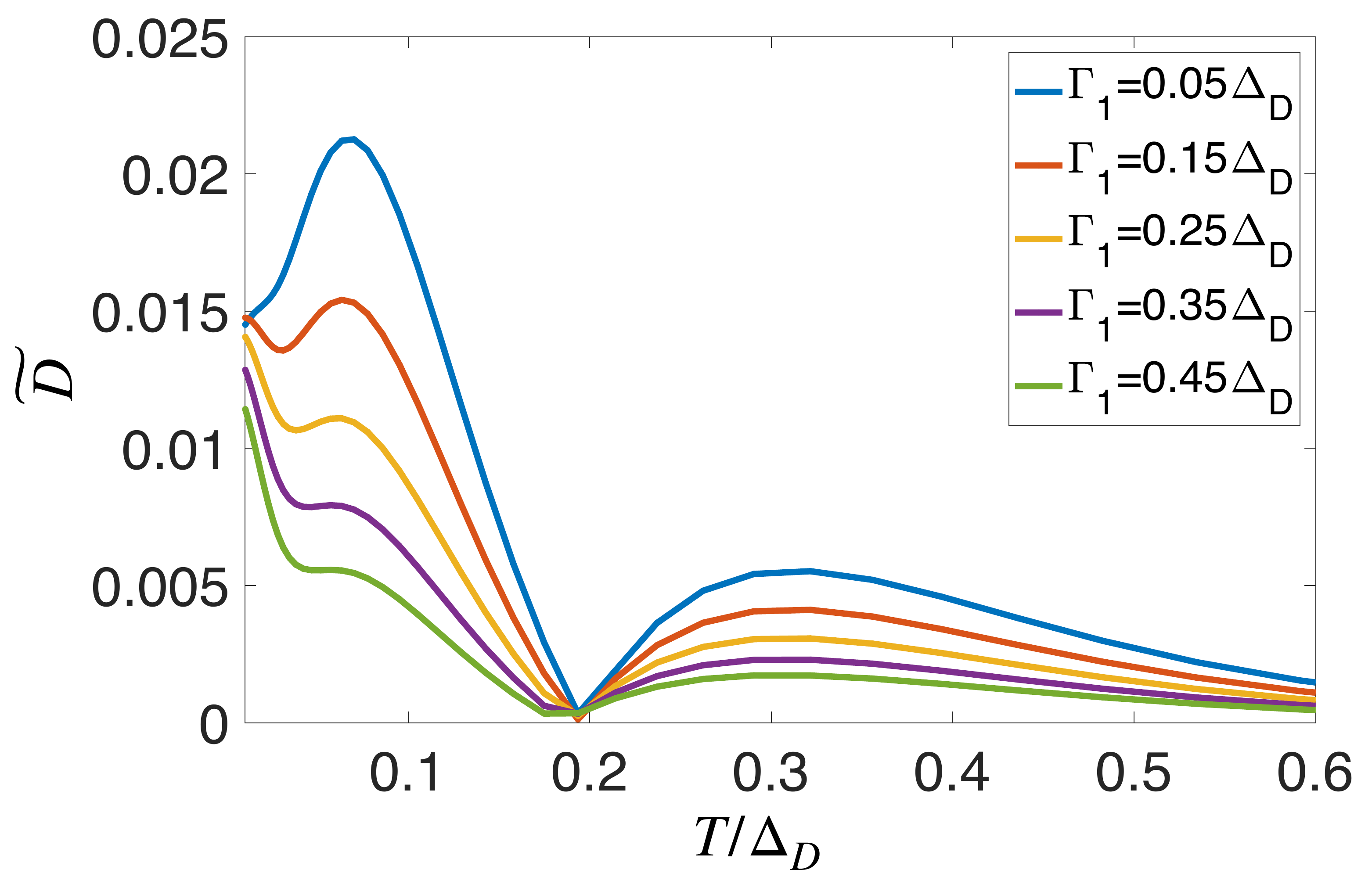}
	}
	
	\caption{{\bf Complex temperature dependence of LE-DOS amplitude:} LE-DOS oscillation amplitude $\widetilde{D}$ at frequency $F_0$ [normalized by $\widetilde{D}_{v=0}(T=0)$ for zero hybridization] as a function of temperature for scattering rates different from the ones in Fig.\ref{fig:A_M_T}(a). Here $\Gamma_2=0.05\Delta_D$, and the results are obtained using the energy eigenvalues $\mathcal{E}_{l,b\pm}$ in Eq.\eqref{eq:LLeigen}.}
	\label{fig:nontriv_amp} 
\end{figure}

 Here we show that the LE-DOS oscillation amplitude at frequency $F_0$ can exhibit more complex temperature dependence at low temperature compared to that in Fig.\ref{fig:A_M_T}(a) for different choices of disorder strengths. In Fig.\ref{fig:nontriv_amp}, we show that for $\Gamma_2=0.05\Delta_D$ and $\Gamma_1=0.15\Delta_D,0.25\Delta_D$, the amplitude $\widetilde{D}$ [normalized by $\widetilde{D}_{v=0}(T=0)$] initially decreases with $T$, following the LK-like form [Eq.\eqref{eq:LEDOS0}] due to in-gap DOS $D_0(T)$, followed by activated increase expected from gap-edge contribution $D_g(T)$ [Eq.\eqref{eq:LEDOSg}]. Also, due to this interplay of $D_0(T)$ and $D_g(T)$, the amplitude can sharply increase at low temperature, as shown for $\Gamma_1=0.35\Delta_D,0.45\Delta_D$.
 Here, with the increase in impurity scattering strength the zero temperature oscillation amplitude does not decrease monotonically, as one expects naively. This can be seen from the $T=0$ oscillation amplitude for $\Gamma_1=0.05\Delta_D,0.15\Delta_D$ in Fig.\ref{fig:nontriv_amp} and $\Gamma\to0$ case shown in Fig.\ref{fig:A_M_T}(a). In this range, $\widetilde{D}$ increases with $\Gamma_1$.
 
\section{Magnetization}\label{sec:S4}
\subsection{Magnetization oscillations at $T=0$} \label{sec:M_zeroT_S}
At zero temperature the magnetization can be obtained from $M=-\partial E(B)/\partial B$, where
\begin{align}
  E(B)&=N_B\sum_l(\mathcal{E}_{l-}-\mu)  
\end{align}
is the total energy, and the chemical potential is inside the gap so that the sum above runs over all energy levels $l$ in the valence band. Here we assume the semiclassical limit and use the energy eigenvalues of Eq.\eqref{eq:LLeigen_Semiclassic}. To see how the oscillations arise,  $\mathcal{E}_{l-}$ can be split into two parts, i.e. $\mathcal{E}_{l-}=\bar{\mathcal{E}}_{l-}+\tilde{\mathcal{E}}_{l-}$, with $\bar{\mathcal{E}}_{l-}=(W+\hbar eBl/m_-)/2$ and
\begin{align}
 \tilde{\mathcal{E}}_{l-}&=-[(W-\hbar eBl/m_+)^2+\Delta_D^2]^{1/2}/2.    
\end{align}
Thus, we can write $E=E_\mathrm{nosc}+E_\mathrm{osc}$ with 
\begin{subequations}
\begin{align}
E_\mathrm{nosc}&=\sum_l[\bar{\mathcal{E}}_{l-}-\mu]\\
E_\mathrm{osc}&=\sum_l\tilde{\mathcal{E}}_{l-}.
\end{align}
\end{subequations}
$E_\mathrm{nosc}$ above is exactly same as that of a completely filled valence band with usual equispaced LLs $\hbar eBl/2m_-$. Hence, $E_\mathrm{nosc}$ cannot give rise to any oscillations and it is a smooth monotonic function of $B$. However, $E_\mathrm{osc}$ corresponds to the total energy due to completely occupied valence band of a particle-hole symmetric band structure [Fig.\ref{fig:schematic}(e,f)], effectively similar to a different model of a hybridization-gap insulator that was considered in ref.\cite{Pal2016}. It was shown there~\cite{Pal2016} that such an insulator exhibits magnetization oscillations. Thus magnetization oscillations arise in our case the same way, albeit from an effective fictitious band structure. We refer the reader to ref.\onlinecite{Pal2016} for a detailed derivation of magnetization oscillations using Euler-MacLaurin expansion for the energy dispersion $\tilde{\mathcal{E}}_\pm(k)$. Here we discuss a simple approximate derivation of the frequency of oscillations. This is further supported by our low-temperature saddle-point calculations discussed in the next sections.

Firstly, it is easy to see that energy levels $\tilde{\mathcal{E}}_{l-}$ periodically crosses the gap edge of the valence band $\mathcal{E}_-(k)$, $\tilde{\mathcal{E}}_v=-\Delta_D/2$ from the hole-like part of the band to the electron-like part [Fig.\ref{fig:schematic}(e,f)] with decreasing field. Here $\tilde{\mathcal{E}}_v$ is obtained from $\partial \tilde{\mathcal{E}}_-(k)/\partial k=0$. Considering two fields $B$ and $B'$ ($B<B'$) such that $\tilde{\mathcal{E}}_{l-}(B')=-\Delta_D/2$ and $\tilde{\mathcal{E}}_{l+1,-}(B)=-\Delta_D/2$, we can find the periodicity
\begin{align}
\frac{1}{B}-\frac{1}{B'}=\frac{\hbar e}{m_+W}=\frac{1}{F_0}
\end{align}
 A simple, albeit heuristic, understanding of how the above $1/B$-periodic crossings affect the total energy can be obtained by neglecting the LLs for $\vert W-\hbar e B l/m_+\vert\lesssim \Delta_D$ and approximating the energy levels as
\begin{align*}
\tilde{\mathcal{E}}_{l-}&\simeq -\frac{1}{2}\left(W-\frac{\hbar e B l}{m_+}\right)-\dots ~~\left(W-\frac{\hbar e B l}{m_+}\right)\gtrsim \Delta_D +\dots\\
&\simeq \frac{1}{2}\left(W-\frac{\hbar e B l}{m_+}\right)-\dots ~~\left(\frac{\hbar e B l}{m_+}-W\right)\gtrsim \Delta_D +\dots
\end{align*}
As a result
\begin{widetext}
\begin{align*}
E_\mathrm{osc}&\approx -\frac{1}{2}\sum_{l\leq m_+B/\hbar e B}\left(W-\frac{\hbar e B l}{m_+}\right)+\frac{1}{2}\sum_{l> m_+B/\hbar e B}\left(W-\frac{\hbar e B l}{m_+}\right)=\sum_{l\leq m_+B/\hbar e B}\left(\frac{\hbar e B l}{m_+}-W\right)+\frac{1}{2}\sum_{l}\left(W-\frac{\hbar e B l}{m_+}\right)
\end{align*}
\end{widetext}
The second term in the last line above is monotonic function of $B$, whereas the first term is an oscillatory function of $1/B$ with frequency $F_0$, exactly like the total energy of a metal with LLs $\hbar e Bl/m_+$ and chemical potential $W$.  Thus, whenever an additional LL enters the electron-like part of the band $\tilde{\mathcal{E}}_-(k)$ from the hole-like part through the gap edge $-\Delta_D/2$, the total energy sharply changes leading to $1/B$-periodic oscillations of the magnetization.

\subsection{Oscillatory part of magnetization in the semiclassical limit}\label{subsec:Mamp}
The grand potential (per unit area) of the model of Eq.\eqref{eq:Model} in the presence of impurity scattering can be written as 
\begin{align}
\Omega(T)&=-TN_B\sum_{\omega_n}\mathrm{Tr}\ln[-\beta \mathbb{G}^{-1}(\imath \omega_n)]e^{\imath \omega_n 0^+}.
\end{align}
Here $\mathbb{G}(\imath \omega_n)$ is the single-particle Green's function matrix in the combined LL index, band and spin space and the `$\mathrm{Tr}$' acts on the same space. For example, in the absence of magnetic field, $\mathbb{G}(\imath \omega_n)$ can be obtained from
\begin{align}
G^{-1}(\bs{k},\imath \omega_n)&=\imath \omega_n+\mu_0-H(\bs{k})-\imath \begin{bmatrix}\Gamma_1 \mathbbm{1} & 0 \\ 0 & \Gamma_2\mathbbm{1}\end{bmatrix}\mathrm{sgn}(\omega_n),
\end{align}
which implies $(\Gamma_1,\Gamma_2)\to (-\Gamma_1,-\Gamma_2)$ for $\omega_n\to -\omega_n$ ($\omega_n>0$). Since, $\mathcal{E}_\pm (k,-\Gamma_1,-\Gamma_2)= \mathcal{E}_\pm^* (k,\Gamma_1,\Gamma_2)$, the Green function in the diagonal basis is $G_\pm(\bs{k},\imath \omega_n)=[\imath \omega_n+\mu_0-\mathcal{E}_{\pm}(k)]^{-1}$ for $\omega_n>0$ and $G_\pm(\bs{k},\imath \omega_n)=[\imath \omega_n+\mu_0-\mathcal{E}^*_{\pm}(k)]^{-1}$ for $\omega_n<0$. Similarly, for $B\neq 0$,  $G_{l,bp}(\imath \omega_n)=[\imath \omega_n+\mu_0-\mathcal{E}_{l,bp}]^{-1}\theta(\omega_n)+[\imath \omega_n+\mu_0-\mathcal{E}^*_{l,bp}]^{-1}\theta(-\omega_n)$. Thus, the grand potential can be written as
\begin{align}
    \Omega(T)&=-TN_B\sum_{lbp,\omega_n>0}\ln(\mathcal{E}_{l,bp}-\mu_0-\imath \omega_n)e^{\imath \omega_n 0^+}+\mathrm{c.c.} \label{eq:GrandPot_S}
\end{align}

In the semi-classic limit($\mu_0\gg\hbar\omega_c$) we replace the eigen energies $\mathcal{E}_{l,bp}$ with $\mathcal{E}_{l\pm}$ [Eq.\ref{eq:LLeigen_Semiclassic}]. We convert the LL summation to an integral using Poisson summation formula and extract the oscillatory component of the $\Omega(T)$ through an integration by parts,
\begin{widetext}
\begin{align*}
\Omega(T)=& 2TN_B\sum_{\omega_n>0}\sum_{k\neq 0} \int_{0^-}^{\infty}dl \frac{e^{2\pi\imath kl}}{2\pi\imath k} \frac{\frac{d}{dl}\left[\left(\mathcal{E}_{+}(l)-\mu_0-\imath\omega_n \right)\left(\mathcal{E}_{-}(l)-\mu_0-\imath\omega_n \right)\right]}{\left[ \mathcal{E}_{+}(l)-\mu_0-\imath\omega_n \right]\left[\mathcal{E}_{-}(l)-\mu_0-\imath\omega_n \right]}  +\mathrm{c.c.}
\end{align*}
 Using the quadratic nature of the function $[ \mathcal{E}_{+}(l)-\mu-\imath\omega_n][\mathcal{E}_{-}(l)-\mu-\imath\omega_n]$, we obtain
\begin{align*}
\Omega(T)\simeq&2TN_B\sum_{\omega_n>0}\sum_{k=1}^{\infty}  \int_{-\infty}^{\infty} dl\frac{e^{2\pi\imath kl}-e^{-2\pi\imath kl}}{2\pi\imath k}\left( \frac{1}{l-l_+(\imath \omega_n)}+\frac{1}{l-l_-(\imath \omega_n)}\right)+\mathrm{c.c.}
\end{align*}
\end{widetext}
It is easy to verify that the poles $l_\pm(n)\equiv l_\pm(\imath\omega_n)$ in the above equation is the same as the poles obtained in Eq.\ref{eq:poles} while calculating DOS, with $\xi\to \imath \omega_n$ in the argument of $l_p(\xi)$. In the above, we have also extended the lower limit of the integral over $l$ to $-\infty$ since $\mathrm{Re}[l_\pm(n)]\gg 1$ in the semiclassical limit. Performing the contour integration over $l$, we obtain
\begin{align*}
\Omega(T)\simeq&2TN_B \sum_{p,k>0,\omega_n>0}\frac{1}{k} e^{2\pi\imath ks_p(n)l_p(n)}+\mathrm{c.c.}
\end{align*}
where, $s_p(n)=\mathrm{sgn}[\mathrm{Im}\{l_p(n)\}]=\mathrm{sgn}[\mathrm{Im}\{x_p(n)\}]$. It can be seen from Eq.\eqref{eq:poles} that $l_p(n)=(F_0/B)+x_p(n)$ and $l^*_\pm(n)=(F_0/B)-x_p(n)$ since $x_p(n)$ is purely imaginary. Thus, we get
\begin{align*}
\Omega(T)&\simeq 4TN_B \sum_{p,k>0,\omega_n>0}\frac{\cos{\left[2\pi k (F_0/B)\right]}}{k} e^{2\pi\imath ks_p(n)x_p(n)}
\end{align*}
We obtain the oscillatory component of the magnetization from $M=-\partial \Omega/\partial B$. The dominant, $\mathcal{O}(\mu_0/\hbar \omega_{c1})$, contribution to magnetization in the semiclassical limit comes from the field derivative of the cosine term in the above equation and is given by
\begin{align}
    M\simeq\frac{8\pi T\mu_0}{\hbar\omega_{c1}\phi_0} \sum_{p, k>0,\omega_n>0} \sin{\left[2\pi k \left(\frac{F_0}{B}\right)\right]}e^{2\pi \imath ks_p(n)x_p(n)}.
    \label{eq:Supp_M_semi}
\end{align}
Based on the low-temperature approximation discussed below it can be shown that the terms neglected above are smaller by factors of $\mathcal{O}(\hbar\omega_c/\mu_0,T/\mu_0,\Gamma/\mu_0,\Delta_I/\mu_0)$.

\subsection{Magnetization oscillations at low temperature}\label{subsec:M_Tpeak}
We rewrite the oscillatory part of magnetization given in Eq.\ref{eq:Supp_M_semi} as
\begin{align}
M\approx & \frac{8\pi\mu_{0}}{\hbar\omega_{c1}\phi_{0}}\left[\sum_{p,k>0} T\sum_{n=0}^{\infty}F_{kp}(n)\right] \label{eq:M_S}
\end{align}
where,
\begin{align*}
F_{kp}(n)= & \sin{\left[2\pi k \left(\frac{F_0}{B}\right)\right]}e^{2\pi \imath ks_p(n)x_p(n)}
\end{align*}

We use the Euler Maclaurin formula $
\sum_{n=a}^{b}f(n)=  \int_{a}^{b}f(x)dx+(1/2)[f(a)+f(b)]+(1/12)[f'(b)-f'(a)]-\dots$
to evaluate the sum at low temperature giving,
\begin{align}
T\sum_{n=0}^{\infty}F_{kp}(n)\approx  T\int_{0}^{\infty}dnF_{kp}(n)+\frac{1}{2}TF_{kp}(0)+\dots\label{eq:EulerMacLaurinS}
\end{align}
where, we use the fact that $F_{kp}(n\to\infty)\to0$. Doing a variable transformation $\omega=(2n+1)\pi T$ we get
\begin{align}
&T\int_{0}^{\infty}dnF_{kp}(n)\nonumber \\
&= \int_{0}^{\infty}\frac{d\omega}{2\pi}\sin{\left[2\pi k  \left(\frac{F_0}{B}\right)\right]}e^{2\pi \imath ks_{p}(\imath \omega)x_{p}(\imath \omega)} \label{eq:omegaIntS}
\end{align}
with 
\begin{align*}
x_{p}(\imath\omega)= (\imath/2\hbar\omega_{c})[m_{r}\omega+\Gamma_{r}+p\sqrt{(\omega+\Gamma_c)^{2}+(\Delta_I/2)^2}].
\end{align*}
The integral in Eq.\eqref{eq:omegaIntS} does not depend on temperature and leads to a constant contribution to magnetization oscillations for $T\to 0$. We again evaluate the above integral by saddle point method. The condition $\partial x_{p}(\omega)/\partial\omega=0$ gives
\begin{align*}
\omega_{\zeta}= & -\Gamma_{c}-\zeta\frac{m_{r}}{\sqrt{1-m_{r}^{2}}}\frac{\Delta_I}{2}
\end{align*}
We only take the saddle point with $\zeta=1$, denoted as $\tilde{\omega}$, which falls on the path
of the integration $\int_{0}^{\infty}d\omega$. This leads to
\begin{align*}
x_{p}(\imath \tilde{\omega})= & \frac{\imath}{2\hbar\omega_{c}}\left[(m_{r}\Gamma_c+\Gamma_{r})+\frac{(p+ m_{r}^{2})\Delta_I}{2\sqrt{1-m_{r}^{2}}}\right],
\end{align*}
and the pole 
\begin{align*}
l_p(\imath \tilde{\omega})&=\frac{F_0}{B}+\frac{\imath}{2\omega_c\tau_p}
\end{align*}
with
\begin{align*}
\frac{1}{\tau_p}&=\frac{1}{\hbar}\left[\frac{2m_+\gamma}{m_1+m_2}+\frac{(p+m_r^2)\Delta_I}{2\sqrt{1-m_r^2}}\right]
\end{align*}
and $s_{p}(\tilde{\omega})= \mathrm{sgn}[\tau_p]$. Since, $m_r<1$, for the limit $\Delta_I\gg \gamma$, $s_p(\tilde{\omega})=p$. To perform the $\omega$ integral in Eq.\eqref{eq:omegaIntS} using the above saddle point, we expand around the saddle point $\omega=\tilde{\omega}+\xi$, i.e.
\begin{align*}
2\pi\imath ks_{p}(\imath \omega)x_{p}(\imath \omega)\approx & 2\pi\imath k p\left[x_{p}(\tilde{\omega})+\frac{1}{2}\left(\frac{\partial^{2}x_{p}}{\partial\omega^{2}}\right)_{\omega=\tilde{\omega}}\xi^{2}\right]
\end{align*}
where,
\begin{align*}
2\pi\imath kp\left(\frac{\partial^{2}x_{p}}{\partial\omega^{2}}\right)_{\omega=\tilde{\omega}}=  -\frac{2\pi k}{\hbar\omega_{c}\Delta_I}(1-m_{r}^{2})^{3/2}
\end{align*}
Now we perform the integral,
\begin{align*}
&\int d\xi\exp\left(-\frac{\pi k}{\hbar\omega_{c}\Delta_I}(1-m_{r}^{2})^{3/2}\xi^{2}\right)\nonumber\\
&=\sqrt{\frac{\hbar\omega_{c}\Delta_I}{k}}(1-m_{r}^{2})^{3/4},
\end{align*}
finally to obtain
\begin{align}
&T\int_{0}^{\infty}dnF_{kp}(n)\nonumber\\
&\simeq  \frac{1}{2\pi}\sqrt{\frac{\hbar\omega_{c}\Delta_I}{k}}(1-m_{r}^2)^{3/4}\sin{\left[2\pi k\left(\frac{F_0}{B}\right)\right]}
  e^{-\pi k/\omega_c\vert\tau_p\vert}. \label{eq:FkpS}
\end{align}
Here it is important to note that, unlike the saddle-point approximation for LE-DOS discussed in Sec.\ref{Sec:DgS}, the saddle-point integral above is only controlled for $m_r\to 1$ i.e. $m_2\gg m_1$, when the Gaussian integrand becomes sharply peaked around the saddle point. The temperature dependence of the magnetization oscillation amplitude
comes from the second and higher order terms in Euler MacLaurin formula
[Eq.\ref{eq:EulerMacLaurin}], namely
\begin{align*}
\frac{1}{2}TF_{pk}(0)= & \frac{T}{2}\sin\left[2\pi k\left(\frac{F_0}{B}\right)\right]e^{2\pi iks_{p}(0)x_{p}(0)}.
\end{align*}
Here, $x_p(0)=\imath/[2\omega_c \tau_{1p}(T)]$ with
\begin{align*}
\frac{1}{\tau_{1p}}= & \frac{1}{\hbar}\left[\sqrt{(\pi T+\Gamma_{c})^{2}+(\Delta_I/2)^2}+p(\Gamma_r-m_r\pi T)\right],
\end{align*}
and $s_{p}(0)= p$ since $\Gamma_{r}\leq\Gamma_{c},~m_{r}<1$. Thus we get
\begin{align}
\frac{1}{2}TF_{kp}(0)= & \frac{T}{2}\sin\left[2\pi k \left(\frac{F_0}{B}\right)\right]e^{-\pi k/\omega_c\tau_{1p}(T)} \label{eq:Fkp_TS}
\end{align}
From $\partial\tau_{1p}^{-1}/\partial T=0$, we find out that $\tau_{1p}^{-1}(T)$ has a minimum at some temperature $T_{peak}$, and hence a peak for oscillation amplitude. This gives
\begin{align*}
T_{peak}= & \frac{1}{\pi}\left(-\Gamma_{c}\pm\frac{\Delta_Im_{r}}{2\sqrt{1-m_{r}^{2}}}\right)
\end{align*}
Thus we see that there could be a peak oscillation amplitude for only
one of the contributions `$+$', i.e.
\begin{align*}
T_{peak}= & \frac{1}{\pi}\left(\frac{1}{2}\frac{m_{2}-m_{1}}{m_{1}+m_{2}}\Delta_{D}-\frac{m_{1}\Gamma_{1}+m_{2}\Gamma_{2}}{m_{1}+m_{2}}\right)
\end{align*}
For $\Delta_I\gg \Gamma$, $T_{peak}\sim\Delta_I/(2\pi)$ and it
moves to lower temperature with increasing $\Gamma_{c}$. Using Eqs.\eqref{eq:EulerMacLaurinS},\eqref{eq:FkpS},\eqref{eq:Fkp_TS} in Eq.\eqref{eq:M_S}, we obtain the expression for magnetization [Eq.\eqref{eq:dHvA}],
\begin{widetext}
\begin{align}
M&\simeq \frac{4\mu_0}{\hbar\omega_{c1}\phi_0}\sqrt{\hbar \omega_c\Delta_I}\sin\left[2\pi k \left(\frac{F_0}{B}\right)\right]\sum_{p,k=1}^\infty  \left[\frac{(1-m_r^2)^{3/2}}{\sqrt{k}}e^{-\pi k/\omega_c\vert \tau_p\vert}+\frac{\pi T}{\sqrt{\hbar \omega_c\Delta_I}}e^{-\pi k/\omega_c\tau_{1p}(T)}\right]
\end{align}
\end{widetext}

\section{Numerical calculation of magnetization} \label{sec:M_num_S}
In our numerical calculations for the disorder-free case $\Gamma=0$, we compute $M(T)$ using
\begin{align}
	\Omega(T)&=-\int_{-\infty}^{\infty}d\xi \frac{\partial n_\mathrm{F}(\xi,T)}{\partial \xi}  \Omega(\xi,T=0), 
\end{align} 
for the grand potential at finite temperature and for the chemical potential $\mu_0$. Here
\begin{align}
 \Omega(\xi,T=0)&=N_B\sum_{\mathcal{E}_{l,bp}\leq \mu_0+\xi}(\mathcal{E}_{l,bp}-\mu_0-\xi)
\end{align}
is the grand potential or total energy at zero temperature for a chemical potential $\mu_0+\xi$. For numerically evaluating the above we put an upper cutoff $\Lambda$ for the LL index $l$. Furthermore, to extract the oscillatory part of the grand potential we subtract from $\Omega(\xi,T=0)$ a large non-oscillatory contribution $2\sum_{l=0}^\Lambda(\epsilon_{2,l}-\mu_0-\xi)$ (factor 2 for the spin degeneracy), which is the grand potential for completely filled valence band in the absence of hybridization. The magnetization is obtained by numerical differentiation of $\Omega(T)$ with respect to $B$. We have verified that results obtained for $M(T)$ are insensitive to the choice of $\Lambda$ for sufficiently large $\Lambda$. The results for magnetization oscillations are shown in Fig.\ref{fig:M_DOS}(b). The amplitude for the Fourier component at frequency $F_0$, $\widetilde{M}(T)$ shown in Fig.\ref{fig:A_M_T}(b) as a function of $T$, is obtained by fast Fourier transform (FFT) of $M(T)$ with respect to $1/B$. We plot the amplitude $\widetilde{M}(T)/\widetilde{M}_{v=0}(0)$ normalized by the $T=0$ value $\widetilde{M}_{v=0}(0)$ for the zero hybridization case.

To evaluate $M(T)$ for $\Gamma\neq 0$, we use the expression for grand potential given in Eq.\eqref{eq:GrandPot_S}, and following steps similar to that discussed in Sec.\ref{subsec:Mamp}
obtain the magnetization amplitude for the $k=1$ harmonic of the fundamental frequency $F_0$, i.e. 
\begin{equation}
	\label{12a}
	\widetilde{M}(T)\simeq\frac{8\pi T\mu_0}{\hbar\omega_{c1}\phi_0} \sum_{bp,\omega_n>0} e^{-2\pi |\mathrm{Im}[l_{b,p}(n)]|}
\end{equation}
where $l_{b\pm}(n)$ ($b=\uparrow\downarrow,\downarrow\uparrow$) are the two poles of the function $[( \mathcal{E}_{b+}(l)-\mu_0-\imath\omega_n )(\mathcal{E}_{b-}(l)-\mu_0-\imath\omega_n) ]^{-1}$. We perform the Matsubara summation above numerically with a cutoff for the largest Matsubara frequency. Note that we use the original energy eigenvalues of Eq.\eqref{eq:LLeigen}, as opposed to the semiclassical eigenvalues [Eq.\eqref{eq:LLeigen_Semiclassic}] that are used in Sec.\ref{subsec:Mamp}.\\\

\section{Transport} \label{sec:Transport}
\begin{figure}[htb!]
	\centering
	{
		\includegraphics[width=0.4\textwidth]{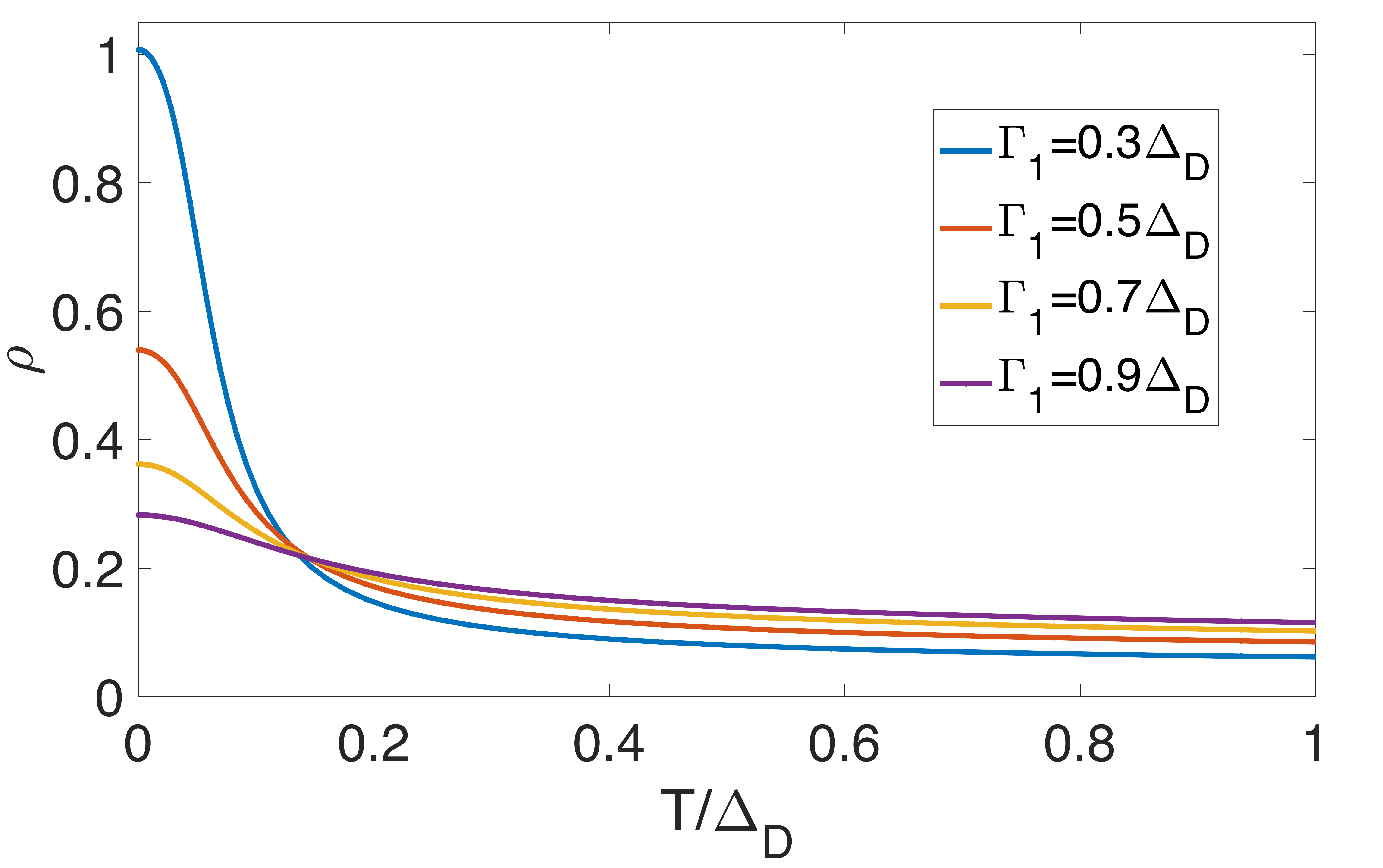}
	}
	
	\caption{{\bf Transport:} Resistivity $\rho(T)$ as a function of temperature for several disorder strengths using Kubo formula for zero magnetic field case. The unit in the $y$-axis for resistivity is $h/e^2$.	Here $\Gamma_2=0.1\Delta_D$.}
	\label{fig:transport} 
\end{figure}
Here we show that even in the presence of disorder, which induces finite DOS inside the gap, the system exhibits $d\rho/dT<0$ at any finite temperature like an insulator. Nevertheless, the system, strictly speaking, remains a metal with finite resistivity $\rho(T=0)$ at zero temperature. To this end, we calculate the conductivity $\sigma$ using the Kubo formula,
\begin{equation}
\label{13a}
\sigma=2e^2\pi\hbar  \int d\omega\left(-\frac{\partial n_\mathrm{F}}{\partial \omega}\right) \sum_{p=\pm} \int \frac{d^2k}{(2\pi)^2} v_{xp}^2(k) A_{p}(k,\omega)^2 
\end{equation}
 Here $A_p(k,\omega)=-(1/\pi)\mathrm{Im}[1/(\omega-\mathcal{E}_p(k))]$ is the spectral function, and we use the real part of the complex eigen energies to calculate the band velocity, i.e., 
\begin{equation*}
    v_{xp}(k)=\frac{\partial\mathrm{Re}\left[ \mathcal{E}_p (k) \right]}{\partial(\hbar k_x)}.
\end{equation*}
We plot the resistivity $\rho=\sigma^{-1}$ as a function of temperature for several $\Gamma_1$ in Fig.\ref{fig:transport}. The insulating-like upturn ($d\rho/dT<0$) with decreasing temperature is evident.  Nevertheless, the resistivity eventually saturates to a finite value as $T\to0$ implying that the system is actually metallic due to impurity-induced in-gap sates.


\bibliography{QOsc.bib}

\vspace{1mm}

\end{document}